\documentclass[12pt]{book}
\usepackage[utf8]{inputenc}
\usepackage{multicol}
\usepackage{caption}
\usepackage{graphicx}
\usepackage{lipsum}
\usepackage{amssymb}
\usepackage{amsmath}
\usepackage{amsfonts}
\usepackage{bbold}
\usepackage{array}
\usepackage{xcolor}
\usepackage{float}
\usepackage{titlesec}
\usepackage{lilyglyphs}
\usepackage{cite}
\usepackage{bookmark}
\usepackage{multirow}
\usepackage{graphicx}
\usepackage{algorithm}
\usepackage{algorithmic}
\usepackage{pifont}
\usepackage{harmony}
\usepackage{roboto}
\usepackage[T1]{fontenc}
\usepackage{musixtex}
\usepackage{epigraph}
\usepackage[sort=use]{glossaries}
\usepackage{listings}
\usepackage{ragged2e}

\usepackage{relsize,etoolbox}
\AtBeginEnvironment{quote}{\smaller}

\newcommand\MyBox[2]{
  \fbox{\lower0.75cm
    \vbox to 1.7cm{\vfil
      \hbox to 1.7cm{\hfil\parbox{1.4cm}{#1\\#2}\hfil}
      \vfil}%
  }%
}

\usepackage[Lenny]{fncychap}

\usepackage{ifoddpage}

\makeatletter
\renewcommand\@chapapp{Chapitre}
\makeatother

\definecolor{mygreen}{rgb}{0.3, 0.7, 0.0}
\hypersetup{colorlinks=true,
            citecolor = black,
            linkcolor = black,
            urlcolor = black}

\titleformat*{\paragraph}{\bfseries}

\newcommand\citesec[1]{section \fcolorbox{red}{white}{\hyperref[sec:#1]{\ref{sec:#1}}}}
\newcommand\citechp[1]{chapter \fcolorbox{red}{white}{\hyperref[sec:#1]{\ref{chp:#1}}}}

\newcommand\citeapd[1]{appendix \fcolorbox{red}{white}{\hyperref[appendix:#1]{\ref{appendix:#1}}}}

\begin{document}

\begin{titlepage}
\centering
\vspace*{-2cm}

\begin{center}
	\begin{minipage}[c]{\linewidth}
    \centering
	\includegraphics[height=1.5cm]{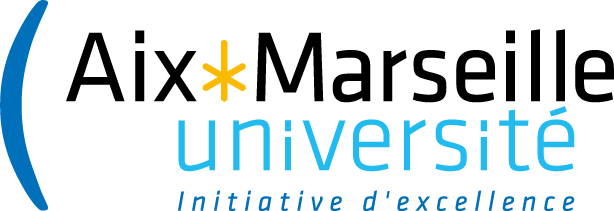}
        \hspace{0.5cm}
        \includegraphics[height=1.5cm]{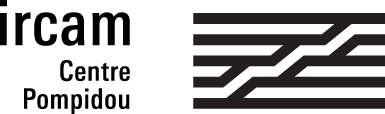}
        \hspace{0.5cm}
        \includegraphics[height=1.5cm]{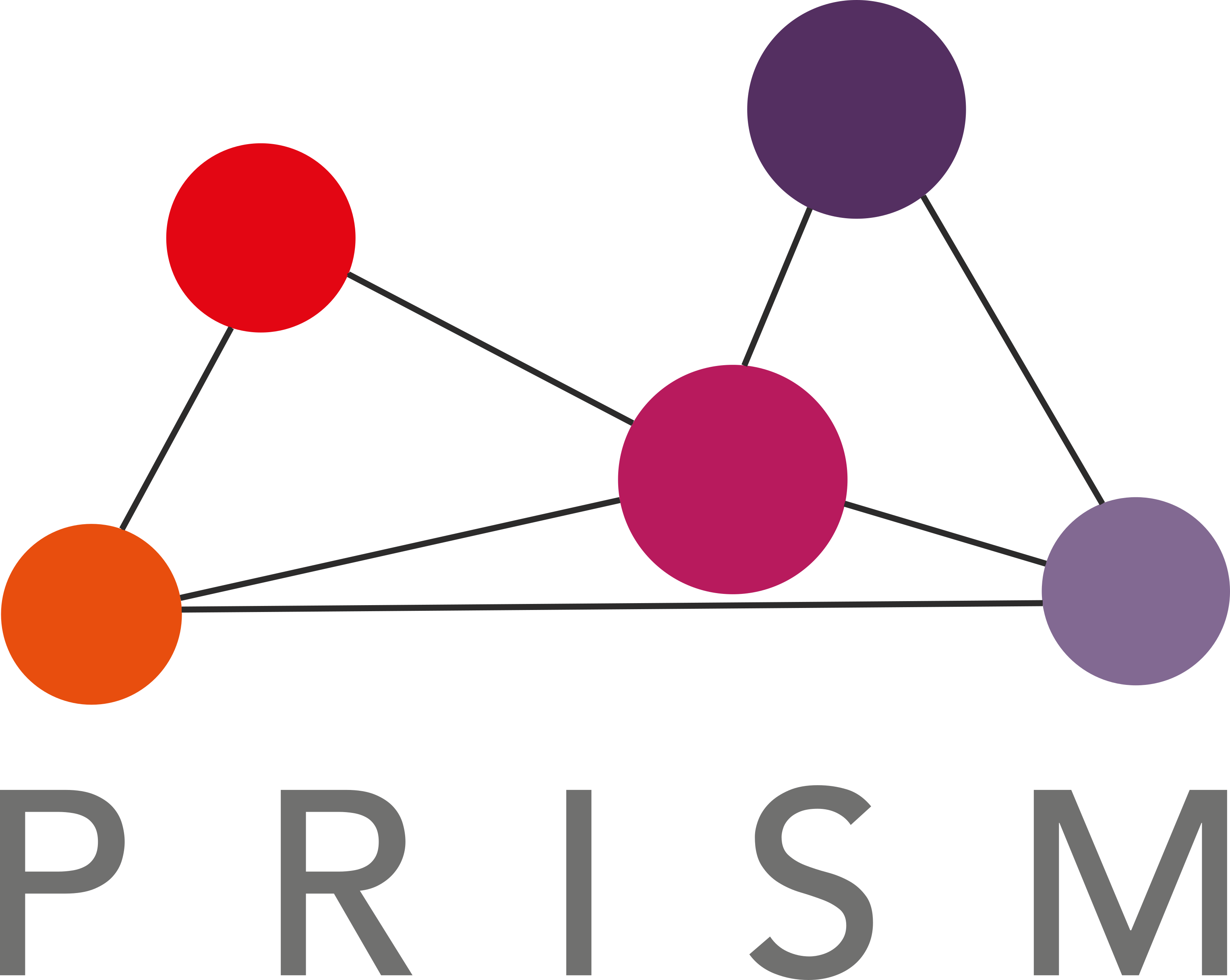}
	\end{minipage}\hfill
\end{center}

\begin{center}
	\vspace{0.4cm}
	\LARGE AIX-MARSEILLE UNIVERSITÉ\\
	\vspace{0.2cm}
	\Large Master Acoustique et Musicologie\\
	\vspace{0.2cm}
	Parcours Musicologie et Création\\
    \begin{center}
		\vspace{1cm}
		Mémoire de recherche-création/développement\\
    \end{center}
	\vspace{0.4cm}
    Discipline: Musicologie\\
    \begin{center}
        \vspace{0.4cm}
        \Large Matéo FAYET\\
        \vspace{0.5cm}
				\LARGE Vivo : une approche multimodale de la synthèse concaténative par corpus dans le cadre d'une oeuvre audiovisuelle immersive\\
				\vspace{0.5cm}
        \vspace{0.8cm}
        \large Sous la direction de Vincent Tiffon (Aix-Marseille Université, PRISM UMR~7061) et la co-direction de Diemo Schwarz (IRCAM STMS, UMR 9912)
    \end{center}
	\vspace{1cm}
    \large Soutenance le 22/06/2023 devant le jury composé de:\\
\end{center}

\vspace{0.2cm} \normalsize
\begin{center}
\begin{tabular}{lll}
    \vspace{0.1cm}
	Antoine GONOT & Examinateur \\
	\vspace{0.1cm}
	Charles DE PAIVA SANTANA & Examinateur \\
    \vspace{0.1cm}
	Vincent TIFFON & Directeur \\
    \vspace{0.1cm}
    Diemo SCHWARZ & Co-Directeur \\
\end{tabular}
\end{center}
\end{titlepage}



\chapter*{Remerciements}
\addcontentsline{toc}{chapter}{Remerciements}
\thispagestyle{empty}
\par Je tiens tout premièrement à remercier Vincent Tiffon et Diemo Schwarz pour leur accompagnement tout au long de ce projet. Leur rigueur scientifique et leur générosité humaine ont été d'une richesse sans précédent et m'ont permis de mener ce projet à terme dans les meilleures conditions possibles. 
\par Merci aussi à Benoît Girerd et son \textit{Briséis} qui nous ont tenus à flots au cours de nos prises de sons et vidéos.
\par Mes remerciements les plus sincères vont aussi à Arthus Touzet, mon ami et collègue, qui m'a apporté son soutien et ses retours les plus précieux tout au long de cette aventure.
\par Merci aussi à tous mes proches qui m'ont soutenu tout au long de ces années d'études qui me permettent aujourd'hui de proposer un projet dont je suis fier. Merci particulièrement à mon frère Solal, pour son soutien inconditionnel et son esprit créatif intarissable, sans qui ce projet n'aurait pu voir le jour. 

\newpage

\vspace*{\fill} 
\begin{quote} 
<<~C’est apparemment là le parachèvement de
l’art pour l’art. L’humanité, qui jadis avec Homère avait été objet de
contemplation pour les dieux olympiens, l’est maintenant devenue pour
elle-même. Son aliénation d’elle-même par elle-même a atteint ce degré
qui lui fait vivre sa propre destruction comme une sensation esthétique de
tout premier ordre.~>>~\cite{benjamin}

\end{quote}
\begin{flushright}
Walter Benjamin
\end{flushright}
\vspace*{\fill}


\tableofcontents


\chapter*{Introduction}
\addcontentsline{toc}{chapter}{Introduction} \markboth{INTRODUCTION}{}
\par L'intégration du son à l'image et de l'image au son a été l'objet de nombreuses recherches et expérimentations au cours du vingtième siècle. Du cinéma expérimental, aux créations musicales immersives et interactives, les applications artistiques sont multiples. Si les détails de ce lien sont encore aujourd'hui discutés par les scientifiques, de nombreuses applications proposent d'explorer les complexités de la perception multimodale autour de développements méthodologiques et artistiques.

\par La synthèse concaténative par corpus permet d'explorer des échantillons sonores classifiés en fonctions de leurs caractéristiques physiques et perceptives. Cet outil est utilisé par de nombreux artistes dans un contexte de création musicale et en design sonore. Aujourd'hui, cette technique est élargie à l'exploration d'images fixes déclenchées parallèlement aux échantillons sonores sélectionnés.

\par La présente recherche--création vise à étendre le concept de synthèse concaténative par corpus audiovisuelle à une utilisation vidéo. Ce document présente les outils développés de mars à juin 2023 utiles à la réalisation d'une oeuvre immersive audiovisuelle en collaboration avec le vidéaste Solal Fayet. Ce mémoire de recherche--création débute dans un premier temps avec un état des connaissances actuel soulevant les enjeux du domaine et de cette recherche ainsi que les créations artistiques associées. Nous étudierons dans un second temps le processus de développement des outils d'analyse d'image (\textit{ViVo}) ainsi que leur fonctionnement et leur modélisation avec les outils de création sonore. En outre, nous nous pencherons sur le développement d'un environnement de VJing et l'adaptation de contrôleurs MIDI. Nous décrirons ensuite les étapes de développement nécessaires au mapping audiovisuel et au VJing ainsi que la mise en commun des outils préexistants avec les nouveaux outils. De plus, nous explorerons les problématiques liées à l'appropriation d'une technologie émergente ainsi que sa diffusion.  Dans un dernier temps, nous analyserons en détail le processus de création réalisée en résidence et présentée à la Fabulerie\footnote{\hspace{0.1cm}\href{https://lafabulerie.com/}{La Fabulerie} (Marseille) est un tiers lieu culturel abritant une fabrique numérique (\textit{fab lab}) qui propose une programmation culturelle musicale et visuelle.}. Dans cette même partie, nous explorerons les techniques mises en oeuvre dans la constitutions des corpus sonores et visuels réalisés. Seront proposées pour cela des méthodes d'analyse musicologiques adaptées aux médiums explorés au cours de cette proposition artistique. Ainsi, l'on pourra étudier les différentes problématiques qui ont émergées au sein du complexe art, science, technologie et proposer des alternatives concrètes en vue d'une évolution des éléments propres à la recherche, à la création et au développement.


\chapter{État des connaissances}

\par Ce chapitre vise à répertorier l'état de l'art actuel concernant les démarches scientifiques qui étudient le lien image/son d'un point de vue perceptif, et les démarches artistiques qui mettent en avant une esthétique visant à générer des sons en lien avec les images dans le but d'un résultat unifié. 

\section{Contexte scientifique}

\subsection{La synthèse concaténative}

\par La synthèse concaténative par corpus (CBCS\footnote{~\textit{Corpus Based Concatenative Synthesis}.}) a été explorée pour la première fois dans un contexte de création musicale par Diemo Schwarz dans son travail de thèse dès 1999 puis appliquée en temps réel avec \textit{CataRT} en 2005~\cite{Schwarz2006}. Ce type de synthèse audionumérique permet de jouer des grains issus d'un corpus de sons segmentés et analysés selon leur proximité à une position cible dans un espace de descripteurs~\cite{schwarz2006real}. Son usage est aujourd'hui étendu à un contexte de performance musicale, de design sonore et d'installations tout en permettant une exploration du corpus interactive de la part des utilisateurs~\cite{Schwarz2006}. \textit{CataRT} est aujourd'hui utilisable sous trois différentes formes : 
\begin{itemize}
    \item Un ensemble de \textit{patchs} \textit{Max} basés sur le \textit{package MuBu} qui permet de manipuler l'outil dans un environnement de programmation visuelle spécialisé dans la création musicale.
    \item \textit{SkataRT} sur le logiciel \textit{Ableton Live} pour une utilisation sur station audionumérique.
    \item \textit{CataRT Standalone Application} pour une utilisation autonome.
\end{itemize}

\par On note que depuis la création de \textit{CataRT}, de nombreux outils basés sur la synthèse concaténative par corpus ont vu le jour. Certains sont utilisables sur \textit{PureData}\footnote{~\textit{PureData} est un environnement de programmation graphique open source similaire à \textit{Max}.} tels que \textit{timbreID} développé par William Brent en 2010~\cite{brent2010timbre} et \textit{earGram} développé en 2013 par Gilberto Bernardes qui propose un environnement de composition qui embarque du \textit{machine learning}~\cite{bernardes2012eargram}. \textit{AudioGuide}, développé en 2010 par Benjamin Hackbarth est un programme conçu en langage Python développé au cours d'une résidence art--science à l'IRCAM~\cite{hackbarth2010audioguide}.

\par Des travaux récents proposent une alternative à \textit{CataRT}. Le projet FluCoMa a pour but de présenter des outils de composition et d'analyse de sons basés sur des environnements de programmations tels que \textit{Max}, \textit{PureData} et \textit{SuperCollider}\footnote{~SuperCollider est un environnement et un langage de programmation pour la synthèse audio en temps réel et la composition algorithmique.}~\cite{Tremblay2021}. Ce projet vise d'après les auteurs à permettre à des musiciens "techno-fluent"~\cite{Tremblay2021} de s'approprier l'outil comme moyen de création musicale au travers de ces logiciels de \textit{creative coding}\footnote{~<<~programmation créative.~>>}. En travaillant avec des musiciens, l'équipe du projet FluCoMa a extrait des buts précis ainsi que des priorités quant au developpement de leurs outils. Les points essentiels sont l'adaptabilité aux logiciels existants, la pédagogie liée à l'apprentissage des librairies et la possibilité de traiter des jeux de données importants utilisables avec les machines de tout un chacun~\cite{Tremblay2021}. Il est mentionné dans ce projet la volonté d'échanger avec des musiciens qui utilisent différents environnements de programmation créative afin de proposer des évolutions cohérentes avec les demandes de musiciens. C'est la raison pour laquelle l'adaptabilité à différents environnement est appuyée dans ce projet.

\par En parallèle, les travaux de Nick Collins explorent la possibilité d'utiliser simultanément des déscripteurs audio et visuels pour la synthèse basée sur les données dans un contexte technique et artistique~\cite{collins2007audiovisual}. Dans ce projet, cinq descripteurs audio et cinq descripteurs vidéo sont proposés. Un descripteur intéressant consiste à analyser le niveau de détail de l'image en appliquant des transformées de Fourier (FFT) aux images. Si les FFT sont plus généralement utilisées pour la transformation et l'export d'images (\textit{e.g.} la compression au format .jpeg), elle est utile dans l'analyse temps réel du contenu d'une vidéo. Ainsi, nous pouvons déterminer dans quelle plage de fréquence se trouve l'information d'une image et sa netteté.

\par
\begin{figure}[htbp]
\centering
\includegraphics[width=0.80\textwidth]{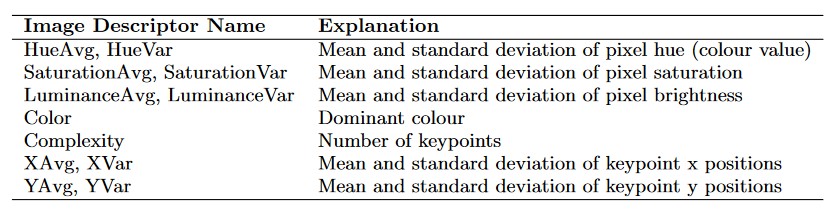}
\caption[Descripteurs vidéo CoCAVS]{\label{cocavs_descriptors} Déscripteurs vidéo du projet CoCAVS \cite{schwarz2023}.}
\end{figure}

 Plus récemment, D. Schwarz propose avec CoCAVS de lier un corpus d'images et un corpus de sons avec l'aide d'un mapping bidimensionnel, d'un algorithme de sélection qui permet de sélectionner les éléments les plus proches imposés par l'un des deux corpus, et des moteurs audio et graphique pour la diffusion des éléments sonores et visuels~\cite{schwarz2023}. Ici, le terme \textit{mapping} désigne une action qui agît sur deux modes à la fois. C'est-à-dire que la sélection d'échantillons des deux corpus est commandée par un seul et même geste. Cette recherche a été réalisée dans le cadre d'une résidence à l'IMéRA en juillet 2022 et une pièce mixte a été présentée dans le planétarium Andromède\footnote{~\href{http://andromede.id.st/}{http://andromede.id.st/}}~\cite{schwarz2023}. Cette recherche a été accompagnée du développement de sept descripteurs vidéo <<~qui caractérisent les qualités perceptives de chaque image par des valeurs numériques~>> \cite{schwarz2023} listés dans la figure \ref{cocavs_descriptors}.

\par Un des éléments essentiels de cette recherche concerne le mapping audiovisuel et l'assignation des descripteurs dans les deux dimensions. L'auteur précise que ces liens consistent plus à mobiliser des analogies intermodales qu'à être fondés sur la perception multiomdale humaine. Par exemple, la brillance d'un son représentée par son centroïde spectral peut être corrélée à la brillance d'une image basée sur l'analyse de la luminance \cite{schwarz2023}. Néanmoins, les descripteurs audio et vidéo peuvent être corrélés dans n'importe quelle configuration afin de laisser une liberté artistique quant aux intentions d'intermodalité. 
\par Un deuxième élément exploré cible le lien de déclenchement entre deux différents corpus. Pour répondre à cela, deux modes de déclenchement sont utilisés. L'un consiste à prendre la position du curseur dans chaque corpus et déclencher les grains les plus proche (\textit{Pre-selection pairing}), l'autre consiste à déclencher le grain d'un corpus après en avoir sélectionné un dans le premier corpus (\textit{Post-selection pairing})~\cite{schwarz2023}. 
\par Cet outil montre que le développement de performances et de créations multimodales est notamment facilité par la conception de descripteurs perceptuellement valides et compréhensifs par les humains dans les différentes dimensions. D. Schwarz évoque par ailleurs que de futurs travaux seraient à mener afin de développer des déscripteurs d'image qui prennent en compte une estimation de la texture, de la saillance des objets, ainsi que les métadonnées\footnote{~L'auteur évoque ici des données enregistrées par l'appareil au moment de la prise de vue : année, heure, focale, vitesse d'obturation, exposition, zoom~\cite{schwarz2023}.}. \par Enfin, l'auteur propose d'étendre ces travaux à des séquences d'images extraites de courtes vidéos~\cite{schwarz2023}. Ainsi <<~CoCAVS explore de nouvelles façons de sonder artistiquement la perception multimodale humaine en créant des associations intermodales dans la navigation gestuelle de corpus communs d'audio et d'images.~>>~\cite{schwarz2023}. Cette recherche ouvre des portes sur les éléments à développer afin de créer un outil pour la création multimodale. Dans cet élan, nous proposerons des outils d'analyse et de mapping qui répondent aux ouvertures de l'auteur, dans une esthétique et une interaction humain-machine similaire~: <<~Enfin, les images ne sont pas générées par une intelligence artificielle, mais choisies par la sensibilité humaine du musicien/interprète audiovisuel. En tant que tel, il offre une approche alternative de l'animation et de la narration, créant une évolution à partir d'instants, et un mouvement à partir de l'immobile.~>>~~\cite{schwarz2023}.

\subsection{Multimodalité audiovisuelle}

\subsubsection{Relation Synesthésique}

\par
\begin{figure}[htbp]
\centering
\includegraphics[width=0.60\textwidth]{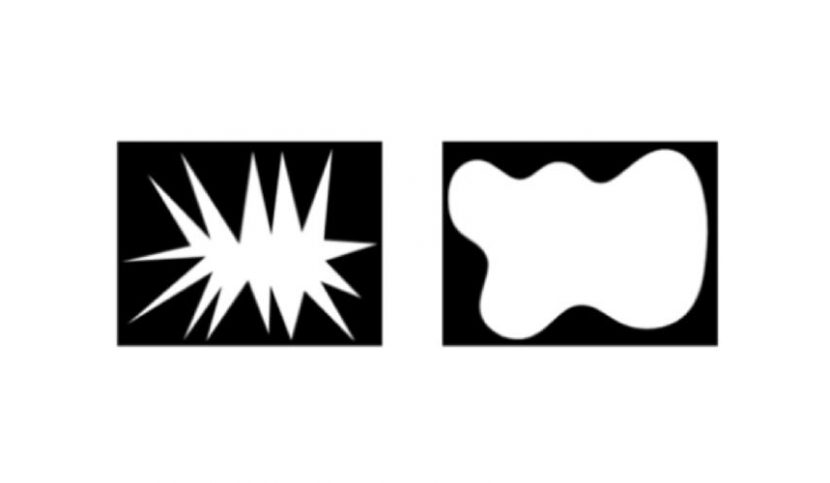}
\caption[Expérience Bouba-Kiki]{\label{bouba_kiki}. Les formes présentées lors de l'expérience Kiki et Bouba \cite{PeifferSmadja2019}.}
\end{figure}

 La perception multisensorielle a fait l'objet de nombreuses recherches depuis le début du vingt-et-unième siècle. Une première approche notable concerne la synésthesie. Ce terme issu de la psychologie fait référence chez Randy Jones et Ben Neville (2005) à un mélange des sens qui parvient chez certains individus \cite{jones}. Lorsqu'une personne perçoit une lettre de l'alphabet d'une certaine couleur, les auteurs décrivent ce phénomène comme ``a true sensory phenomenon rather than a conceptuel one'' \footnote{~<<~Un réel phénomène sensoriel plutôt qu'un phénomène conceptuel [traduction personnelle]~>>.}. Cependant, toujours d'après R. Jones et B. Neville, on observe un phénomène plus généralisable qui a été expérimenté par Wolfgang Kohler, psychologue de la \textit{Gestalttheorie}\footnote{~Psychologie de la forme.} en 1929. Dans cette experience, il est demandé aux participants qui observent la figure \ref{bouba_kiki} ``Which of these is a 'bouba' and which is the 'kiki'?'', ce à quoi quatre-vingt dix pourcent des participants ont répondu que la forme ronde était 'bouba' et que la forme pointue était 'kiki' \cite{jones}. 

Le phénomène Bouba-Kiki est associé par Ramachandran et Hubbard en 2001 \cite{ramachandran2001synaesthesia} à la création d'une connectivité multimodale dans le cerveau ce qui fait de chaque humain, dans une certaine mesure, un synesthète\cite{jones}. Se pose ainsi la question suivante : la synesthésie est-elle un phénomène perceptif ou cognitif~? Après avoir mené cinq expériences à propos de la relation graphème-couleur\footnote{~Le graphème est la plus petite entité d'un système d'écriture~: il désigne une lettre qui transcrit un phonème.} sur des sujets synesthètes, les auteurs déduisent que les couleurs induites sont de nature sensorielle. Ils évoquent cependant le fait qu'il existe différents niveaux de synesthésie chez les individu concernés, et que chaque cas n'est pas nécessairement généralisable. On extrait tout de même une définition de cette approche cognitive du phénomène~: ``Synaesthesia is a curious condition in which an otherwise normal person experiences sensations in one modality when a second modality is stimulated''\footnote{~<<~La synesthésie est un état curieux dans lequel une personne par ailleurs normale expérimente des sensations dans une modalité lorsqu'une autre est stimulée [traduction personnelle]~>>.}. Dans notre cas, il est intéressant de souligner que le but n'est pas de développer un outil exclusivement synesthèsique et cognitif, mais de produire des congruences multimodales perceptives dans le cadre d'oeuvres artistiques.

\subsubsection{Relation timbre-sémantique}

\par William Gaver soumet en 1993 une approche écologique de la perception auditive \cite{gaver1993world} basée sur les travaux de Gibson de perceptions sensorielle globales en 1966 \cite{gibson1966senses} et visuelles en 1979 \cite{gibson1979ecological}. D'après les auteurs, ``perception is usually of complex events and entities in the everyday world''~\cite{gibson1979ecological}~\footnote{~<<~La perception est globalement celle d'événements complexes et d'entités du monde de tous les jours [traduction personnelle]~>>.}. W. Gaver ajoute même que la perception ``is direct, unmediated by inference or memory'' \footnote{~<<~Elle est directe et non générée par l'intermédiaire de l'inférence ou de la mémoire [traduction personnelle]~>>} Cela pose d'après W. Gaver deux questions : qu'est ce que l'on entend ? Et comment l'entendons nous ?.

\begin{figure}[htbp]
\centering
\includegraphics[width=0.60\textwidth]{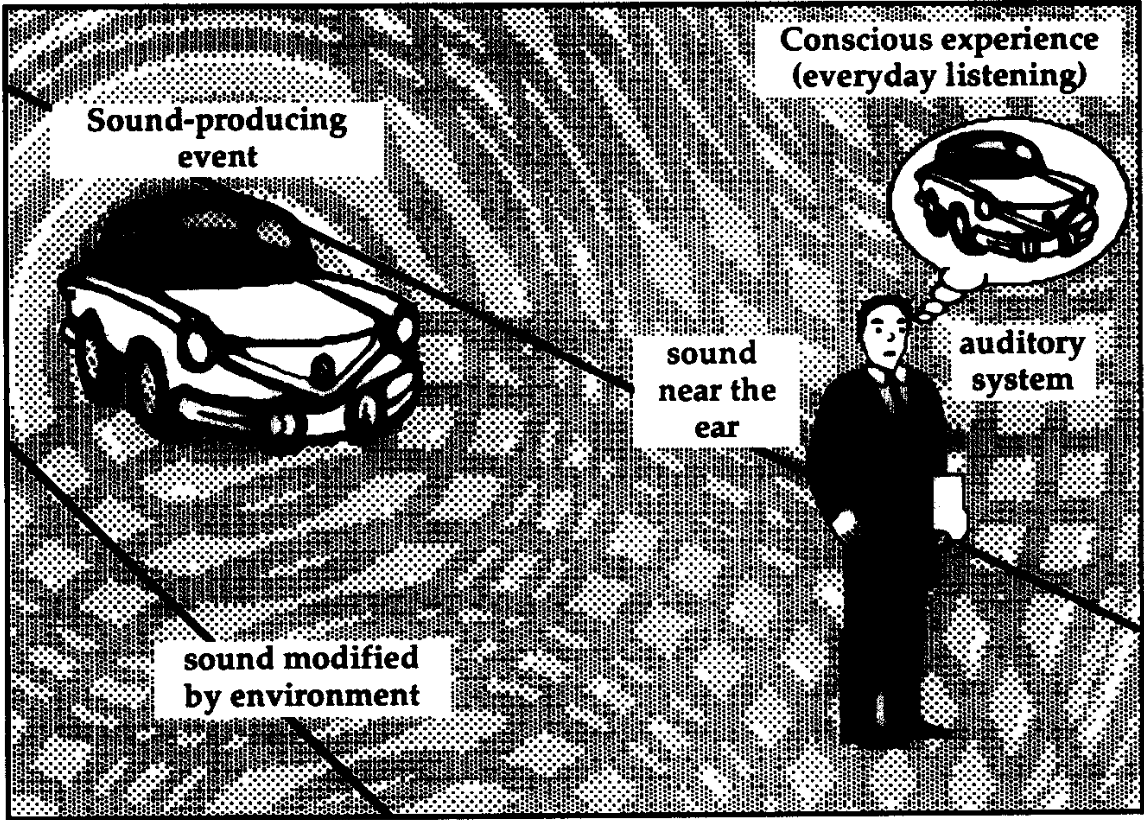}
\caption[Schéma Gaver]{\label{gaver-envi}Principe d'influence du contexte d'écoute sur la perception d'événements sonore \cite{gaver1993world}.}
\end{figure}

\par Par ailleurs, l'équipe Perception et Design Sonore de l'IRCAM propose en 2020 une étude sur la signification sémantique de quatre attributs sonores : brillant, rugueux, rond et chaud \cite{pds}. L'expérience permet de déterminer quelles caractéristiques physiques des sons semblent importantes lors de descriptions sémantiques timbrales de la part de professionnels du son (ingénieurs, designers sonores, musiciens, compositeurs). Si cette étude ne permet pas d'établir de lien entre perception sonore et perception visuelle, elle permet de préciser sur quels paramètres physiques sonores se base la description sémantique de certains sons. Cette récente publication basée sur des études qui permettent de caractériser des sémantiques récurrentes dans les descriptions de certains instruments de musique spécifiques tels que l'orgue \cite{disley2004spectral} et la guitare \cite{traube2004interdisciplinary} généralise ainsi un vocabulaire et une sémantique globale concernant les caractéristiques physiques des sons. La recherche de l'équipe Perception et Design Sonore constitue un véritable point d'entrée dans la conception de notre dispositif multimodal. La possibilité d'extraction de descripteurs sémantiques relatifs aux propriétés physiques des sons nous permet d'avoir un objectif précis sur les paramètres sonores que nous voulons contrôler.

\subsubsection{Définitions}

\par La modalité est assimilée dans divers domaines scientifiques -- comme la psychologie ou la linguistique -- à la technique d'interaction humain-machine à différents niveaux d'abstraction (physique, perceptif, cognitif).~\cite{nigay1996espaces}. En psychologie, le terme désigne une des grandes catégories sensorielles que sont la vision, l'odorat, le goût, l'ouïe et le toucher. Complété du préfixe \textit{multi} qui désigne la notion de nombre, on comprend que l'expression désigne une interaction humain-machine constituée de plusieurs éléments des différentes catégories d'abstraction. Chez les psychologues, <<~la multimodalité d’un système tient au fait que la machine sollicite les capacités multi-sensorielles et cognitives de l’utilisateur~>>~\cite{nigay1996espaces}. Cette notion est étroitement liée avec le concept informatique de \textit{multimédia}. En effet, cette expression chez les professionnels de ce secteur se caractérise <<~en entrée, par l'utilisation simultanée de différents dispositifs physiques, par la gestion en parallèle des événements qui en découlent, et par la construction d'événements de haut niveau d'abstraction à partir d'événements de plus bas niveau [...] en sortie, par différents canaux en parallèle (texte, vidéo etc.)~>>~\cite{nigay1996espaces}. 

\par Nous souhaitons ici proposer une expérience multimodale à l'aide d'outils multimédias. Quels sont alors les différents modes mis en jeux lors de la perception d'évènements audiovisuels et comment doit-on mettre en place la relation entre les différents médias~?

\section{Contexte artistique}

\par Cette section propose de donner un aperçu des oeuvres essentielles de la musique éléctroacoustique dans le cadre d'une composition immersive audiovisuelle mettant en jeu les techniques de synthèse granulaire et de vidéomusique. En lien avec la section précédente, nous ferons émerger des corrélations arts--sciences--technologies nécessaires à la conception d'un outil de création artistique numérique, dans le but de constituer un corpus complet.

\subsection{La synthèse granulaire}

\par Dans le contexte de cette recherche--création, la synthèse concaténative par corpus permet de contrôler en partie un programme de synthèse granulaire. Cette technique est introduite mathématiquement en 1947 par Dennis Gabor \cite{GABOR1947} qui suppose que toute représentation granulaire -- qu'il nomme \textit{quanta acoustique} -- permet de décrire n'importe quel son. D.~Gabor théorise le principe selon lequel un signal peut être décrit par la somme des événements temporels infinitésimaux qui le compose. Il se positionne ainsi en rupture avec le paradogme de la décomposition de Fourier. C'est en 1971 que Iannis Xenakis propose la première théorie de composition granulaire analogique \cite{xenakis1992formalized}. 

\par
\begin{figure}[htbp]
\centering
\includegraphics[width=0.60\textwidth]{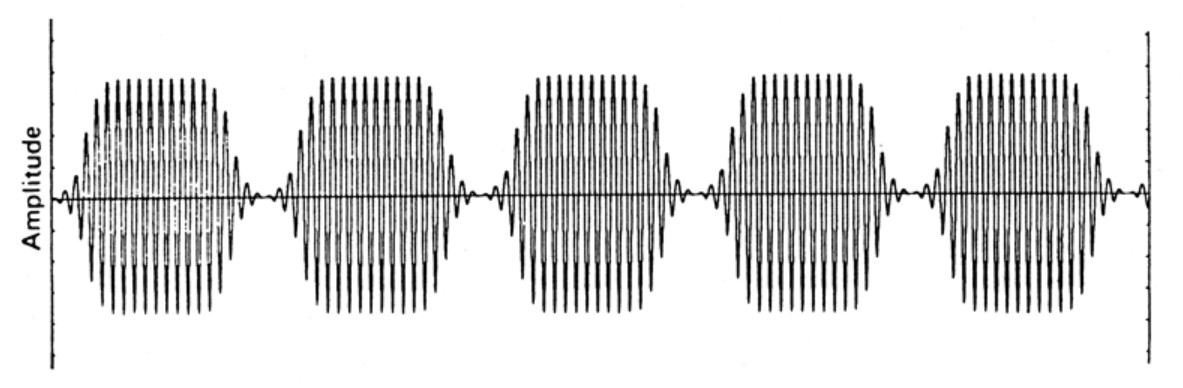}
\caption[Grain-enveloppe]{\label{grain-envelope}Exemple de nuage de grains à durée et enveloppe constante.\footnotemark}
\end{figure}

 Dans son introduction à la synthèse granulaire de 1988 \cite{Roads1988}, Curtis Roads formalise le traitement numérique nécessaire pour parvenir à la granulation des sons sur ordinateur~\footnote{~Si le texe date de 1988, les implémentations numériques de C.~Roads datent de 1975 et 1981.}. En 1988, Barry Truax se propose de décrire le processus de création d'oeuvres temps réel basées sur la synthèse granulaire numérique \cite{Truax1988}. En effet, d'après l'auteur, la composition à l'aide d'outils numériques de synthèse granulaire nécessite de revisiter le mode d'approche de la création. La spatialisation ainsi que le traitement d'une quantité de données relativement élevée par un ordinateur fait émerger de nouvelles problématiques liées à la capacité de calcul des machines. Effectivement, d'après I.~Xennakis~\cite{xenakis1992formalized} il faut générer des grains toutes les vingt à trente millisecondes afin d'obtenir des sons complexes (un grain déclenché toutes les vingt millisecondes correspond à une fréquence audible de cinquante Herz). Cela soulève un problème à mettre en lien avec les textes précédemment cités qui traitent de la relation timbre-sémantique. Si nous savons modifier les paramètres d'un grain pour que sa perception par l'auditeur soit contrôlée, comment faire en sorte que la texture, le nuage ainsi perçu possède aussi ces propriétés sémantiquement descriptibles~? Nous savons en effet désormais qu'au-delà d'une certaine fréquence de déclenchement, l'humain ne percevra plus les grains de façon isolée mais comme un évènement global. B.~Truax évoque donc l'importance des enveloppes de chaque grain dans la perception globale des nuages générés \cite{Truax1988}. Bien que ses algorithmes ne permettent pas de faire évoluer une enveloppe au cours d'une pièce, le choix du temps d'attaque et de relâchement d'un grain (voir figure \ref{grain-envelope}) a une influence non négligeable sur la perception de l'ensemble. Aujourd'hui, les programmes de synthèse granulaire proposent bien entendu de faire évoluer les enveloppes de façon fluide et certains algorithmes tels que celui dans l'\textit{external Max} du GMEM Microsound Universe\footnote{~Le GMU est un \textit{package Max} développé par le pôle recherche du Groupe de Musique Expérimental de Marseille en 2005.} permettent même de charger l'enveloppe sous forme de donnée audio, ce qui permet d'obtenir des enveloppes originales à partir de n'importe quel fichier son~\cite{bascou2005gmu}.

\footnotetext{\href{https://www.sfu.ca/}{https://www.sfu.ca/}}
\par Toutes ces problématiques liées à l'apparition d'un nouveau médium de création musicale sont abordées par les compositeurs et développeurs sous le spectre de ce que l'on appelle la lutherie numérique \cite{battier2002phonographie}. Ce terme emprunté à Marc Battier se trouve employé dans un contexte d'opposition aux outils de synthèse analogique (comme ceux présentés ci-avant dans le cas de I.~Xennakis). Il exprime que le cas de la synthèse analogique est comparable aux instruments acoustiques dans le sens où <<~la matière est le résultat de l'excitation du corps sonore~>> \cite{battier2002phonographie}. En revanche, la synthèse numérique du signal fait émerger des degrés de libertés\footnote{~<<~Les degrés de liberté sont le moyen par lequel l'instrumentiste façonne la matière~>> \cite{battier2002phonographie}.} autrement interprétables. 

\par D'après M.~Battier, cela est notamment dû à la définition instable du concept lui-même qui renvoie à trois notions distinctes~:

\begin{itemize}
    \item L'instrument au sens de corps sonore.
    \item La machine comme moyen de calcul et d'opération programmée.
    \item La représentation qui permet aux deux premières catégories d'interagir et au musicien de spécifier ou de modifier les données circulant dans la lutherie. \cite{battier2002phonographie} 
\end{itemize}

\par Tout en pointant les particularités de la synthèse numérique, l'auteur souligne les bénéfices universels d'un tel développement~: <<~l'informatique crée des langages et des codes qui peuvent être acquis par tous, ce qui offre une plateforme supplémentaire de communication~>>~\cite{battier2002phonographie}. Ce langage de communication dont parle ici M.~Battier est essentiel dans le contrôle des paramètres de la synthèse granulaire. Par essence, la modification micro-temporelle de grains dans le but d'un résultat entier, fait échapper au musicien le contrôle direct de la perception du son obtenu. En d'autres termes, les paramètres de contrôle micro-temporels de la synthèse granulaire ne contrôlent qu'indirectement le nuage finalement obtenu. C'est pourquoi des outils d'analyse et de classification de grains tels que les algorithmes \textit{MuBu} utilisés par \textit{CataRT} paraissent essentiels à la compréhension et à la maîtrise du matériau sonore en question. Conséquemment, nous utiliserons \textit{mubu.concat} comme moteur de synthèse granulaire dans notre pièce éléctroacoustique.

\subsection{Vidéomusique et \textit{Visual Music}}

\subsubsection{Musique concrète}

\par Dans le \textit{Traité des Objets Musicaux} de 1966~\cite{schaeffer_1966}, Pierre Schaeffer présente les différents aspects du concept d'\textit{objet musical}. A la fois par les disciplines de la physique et de la philosophie, l'auteur propose une méthode musicale qui renouvelle le solfège traditionnel afin d'étudier les musiques sous une approche plus juste. Souvent considéré comme un ouvrage interdisciplinaire majeur du vingtième siècle, le \textit{Traité des Objets Musicaux} consitue un socle pour la composition et l'analyse d'oeuvres éléctroacoustiques. De plus, les travaux de l'auteur s'étendent à l'oeuvre audiovisuelle, notamment grâce à la création du Groupe de Recherche Image de l'O.R.T.F. Cette approche interdisciplinaire permet d'établir des corrélations entre un signal physique et l'objet musical, ce qui apparaît comme essentiel lorsque l'on utilise des outils de synthèse sonore numériques.

\par La même année que \textit{L'Etude aux objets}~\cite{etudeobjets} de Pierre Schaeffer, pièce emblématique de la musique concrète, Iannis Xenakis compose \textit{Concret PH}~\cite{concretph}, une pièce qui vise à être diffusée dans le Pavillon Philips\footnote{~Bâtiment conçu par Le Corbusier} à l'exposition internationale de Bruxelles. Si cette pièce électroacoustique n'est pas la première à penser l'espace, elle marque un tournant dans l'histoire de la conception de spatialisation musicale. Bien que les différentes sources existantes ne permettent pas de donner le nombre exact de haut-parleurs réquisitionnés pour l'occasion, on l'estime à quatre-cent~\cite{meric}. Proche collaborateur de Le Corbusier, I.~Xenakis a participé lui même au dessin du bâtiment éphémère\footnote{~Le bâtiment a été détruit un peu moins d'un an après sa construction.}, et l'on peut considérer le complexe architecture/composition comme une conception globale. Travaillant sur de courts matériaux sonores enregistrés sur bande, le compositeur propose une approche granulaire de la création musicale sur support analogique~: 

\begin{quote}
<<~[...] le matériau et la technique de composition utilisés par I.~Xenakis permettaient difficilement un travail sur la forme dynamique temporelle, un travail sur des "entités sonores". En effet, I.~Xenakis pour Concret PH, n’a utilisé que de très courts échantillons de bande extraits d’un enregistrement de braises de charbon de bois en train de se consumer. Seuls les crépitements (dont la durée s’établit entre quelques millièmes et quelques centièmes de seconde) ont été prélevés et subtilement ré-assemblés, selon une méthode annonçant la synthèse granulaire.~>>~\cite{meric}
\end{quote}

Le spectacle accompagné par la suite de \textit{Poème électronique} de Edgard Varèse, était accompagné de <<~lumières colorées mouvantes~>>~\cite{meric}. Si aucun lien direct entre le son et la lumière n'est avéré, nous pouvons considérer cette oeuvre comme une approche à la démarche d'une immersion réfléchie d'un public dans un lieu diffusant des sons spatialisés et lumières contrôlées.

\par Semblablement, les oeuvres artistiques de musique visuelle présentent depuis le début du vingtième siècle différents processus de composition. L'ère de la composition numérique permet entre autres aux artistes d'explorer le domaine de la microtemporalité et d'accéder ainsi à des échelles musicales inférieures au jeu instrumental traditionnel~\cite{billon}. On note de fait un certain nombre de créations audiovisuelles qui synchronisent le rythme sonore aux éléments visuels. En effet très proche du cinéma expérimental (surtout à ses débuts), la musique visuelle soulève artistiquement des problématiques cinématographiques de synchronisation image/son.


\subsubsection{\textit{Visual music}}

\par De son côté, Brian Evans propose en 2005~\cite{evans} un travail préparatoire sur les fondements de la théorie de la \textit{visual music} à travers une analyse géométrique et colorimétrique des images. Cette esthétique propre au cinéma expérimental explore les relations entre les images et le son autour de différentes techniques de créations. Le travail de B.~Evans est accompagné d'une brève histoire de la \textit{visual music}, de la découverte du feu avec les jeux d'ombres sur les murs de grottes, jusqu'aux orgues de couleur qui permettent d'accompagner la musique de jeux de couleurs. En outre, l'auteur extrait de certaines figures visuelles des comportements de créations similaires à la composition musicale. Des concepts tels que la tension et la résolution ainsi que leurs degrés respectifs sont abordés. L'auteur précise que ``the perception of the color is inexact, culturally influenced, and personnal''\footnote{~<<~La perception de la couleur est inexacte, culturellement influencée et personnelle~[traduction~personnelle]>>}~\cite{evans}. Cependant, certaines mesures numériques permettent d'analyser la colorimétrie au travers d'aspects généralisables tels que la luminosité et la saturation (qui sont liées aux valeurs ARGB\footnote{~\textit{Alpha, Red, Green, Blue} (opacité, rouge, vert, bleu), les quatre paramètres sur lequel est basée le traitement d'image numérique.}) \cite{evans}. L'auteur cite entre autres Norman McLaren comme un des acteurs principaux de la \textit{visual music}, dont les oeuvres permettent de comprendre les enjeux du style musical et cinématographique.

\par Les oeuvres de Norman McLaren \textit{Lignes verticales}~\cite{linesverti} et \textit{Lignes horizontales}~\cite{lineshori} citées par Brian Evans proposent dès les années 1960 une première approche artistique de la relation image/son et de la perception multimodale dans un contexte de création artistique. \textit{Lignes horizontales} (1961) présente une accumulation progressive visuelle de lignes horizontales mouvantes. La quantité de lignes augmente avec la complexité sonore tout au long de la pièce. Par complexité, nous entendons ici une augmentation du nombre d'instruments et du nombre de notes que ces derniers jouent. Le support visuel a été créé en manipulant la bande sonore avec une tireuse optique, matériel utilisé à cette époque par les producteurs pour manipuler plusieurs bandes et créer des effets spéciaux~\cite{jones}. Ainsi, N.~McLaren a obtenu plusieurs films de différentes couleurs parfaitement synchronisés avec l'audio. Puisque chaque événement a été traité image par image, chaque élément musical est fortement corrélé à un événement visuel~\cite{jones}. 

\par Il est intéressant de mettre en lien la technique de N.~McLaren avec les procédés d'écriture contrapuntiques de l'époque baroque. Le contrepoint repose en effet sur la superposition de lignes mélodiques (visibles sur la partition), qui s'agrémentent au fur et à mesure du temps afin de faire émerger des mélodies secondaires dites horizontales et des harmonies dites verticales. Ainsi, les voix changent de lignes au cours du morceau créant confusion chez l'auditeur qui écoute la musique sans nécessairement lire la partition. L'oeuvre de Norman McLaren propose de représenter visuellement l'apparition de nouvelles voix, créant chez l'auditeur un fort effet synesthésique \cite{jones}. 

\par On note par ailleurs dans son court-métrage d'animation de 1971 \textit{Synchromy}~\cite{synchromy}, une forte relation image/son liée à l'intensité. Plus un élément sonore est fort, plus le visuel présenté est grand. Dans cette même pièce, on constate des changements de couleurs distincts en fonction des différentes parties présentées, mais aucune logique sonore relative à la couleur ne semble émerger de ce paramètre. 

\par On relève donc dans cette esthétique différents liens image/son. A la différence d'une partition, les visuels ont été créés après composition et enregistrement de l'audio et n'ont servi en aucun cas de support aux artistes. En outre, les éléments ne relèvent d'aucune analyse sonore ou visuelle scientifique liant l'image et le son. La complexité est perçue comme logique, laissant notre perception associer un élément visuel à un élément sonore (\textit{e.g.} un instrument à une ligne). Le lien entre intensité sonore et taille d'élément visuel ne se réfère qu'à la conception socialement admise qu'un son fort est lié à un élément plus grand~: cela correspond-il à une approche artistique de la perception multisensorielle~?

\par Si les fondements et les intentions de la \textit{visual music} semblent relever de problématiques similaires à notre projet, l'esthétique ainsi que les outils mis en jeu semblent différents du projet que nous souhaitons mener. 

\subsubsection{Vidéomusique}

\par Plus récemment, la thèse de Jean-Pierre Moreau publiée en 2018~\cite{moreau2018perception} étudie l'histoire et la réception de la vidéomusique (courant esthétique différent de la \textit{visual music}) et propose une méthode d'analyse des oeuvres vidéomusicales. Son travail permet entre autres de constituer un corpus lexical musicologique autour de cette pratique artistique. J.-P. Moreau évoque ainsi Michel Chion, membre du GRM créé par Pierre Schaeffer cité précédemment et Jean Piché que nous évoquerons ci-après pour les oeuvres vidéomusicales précurseures. 

\par C'est en 1990 que le néologisme \textit{vidéomusique} apparaît chez Jean Piché (en rupture avec la \textit{visual music}) pour ses besoins artistiques~\cite{moreau2018perception}. C'est d'ailleurs dans ses publications dans différents numéros de la revue \textit{Circuit}, que J. Piché développe ses approches de la vidéomusique. Il présente ainsi un langage qui permet d'articuler une cohérence dans la musique éléctroacoustique classique : <<~Reste donc le langage et l'articulation de cette matière dans un discours qu'on espère cohérent~>>~\cite{Pich2010}. D'après lui, <<~Le langage de l'objet sonore (dans le sens schaefferien), sur lequel s'est echafaudée toute la pratique électroacoustique depuis les années 1950, refuse de se plier à une méthode d'analyse cohérente~>>~\cite{Pich2010}. Il soulève que  peu de personnes qui ne participent pas à la production de cette musique possèdent les outils nécessaires à la compréhension de cet objet : <<~Le code est privé~>>~\cite{Pich2010}. L'implication du public dans la réception de l'oeuvre en serait ainsi abîmée. C'est donc, d'après l'auteur, l'expérience de la synchronisation de l'image et du son qui joueraient un rôle intégrateur dans <<~l'expérience esthétique médiatisée~>>. J.~Piché utilise ici le terme \textit{synchrèse} évoqué par Michel Chion en 1991 dans son ouvrage \textit{Audio-Vision}~\cite{chion1993audiovision}. D'après ce dernier, la synchronisation d'un élément visuel et d'un élément auditif est essentiel pour percevoir ces éléments comme un seul. En revanche, daprès J. Piché, la présence permanente de synchrèse nuirait au discours en écrasant entre autres la place de la métaphore. De fait, cette approche peut s'avérer contre-productive. Dans son écrit, J. Piché décrit la \textit{visual music} comme un courant qui vise à représenter un phénomène musical par une visualisation abstraite, ce qui d'après lui, entraîne cet art dans une forme de purisme et d'ésotérisme qu'il compare aux premières expérimentations de musiques électroniques~\cite{Pich2010}.

\par Jean Piché, est aussi l'auteur de nombreuses pièces de vidéomusique. Son approche numérique, permet de faire des montages entre différents éléments visuels que l'on identifie notamment grâce à l'apparition d'événements sonores. Les éléments se révèlent de façon continue et il n'existe pas toujours un lien apparent entre le son et l'image. La pièce \textit{Sièves}~\cite{sieves} présente en 2004 une musique aux évolutions lentes et progressives. C'est entre autres ce lien que l'on trouve entre l'image et le son. Contrairement aux pièces de N. McLaren, il n'y a pas d'équivalence directe entre chaque événement multimodal. Par exemple, une forme plus grande ne révèle pas forcement l'apparition d'un son plus fort. C'est là que nous pouvons différencier la vidéomusique et la \textit{visual music}.

\section{Choix des outils technologiques}

\par Au vu des éléments cités précédemment, nous avons choisi de développer quatre modules d'analyse vidéo (\textit{ViVo}) sur le logiciel \textit{Max/MSP/Jitter}. En effet, \textit{CoCAVS} et \textit{CataRT} sont développés en partie sur \textit{Max/MSP}. De plus, la déclinaison \textit{SkataRT} qui permet d'utiliser un explorateur de corpus sur \textit{Ableton} avec un instrument Max for Live offre de nombreuses possibilités. Si \textit{ViVo} est voué à être utilisable pour différents types d'utilisations et de plateformes, son développement et le test de sa compatibilité avec \textit{CataRT} est une étape essentielle au développement de l'outil. Les possibilités de communication multiplateformes seront offertes par la gestion du protocole OSC sur \textit{Max}. Le protocole OSC a été développé dans le but d'offrir des capacités de communication interlogiciels de façon fiable, précise, simple et avec une gestion généralisable de la sémantique et des données~\cite{wright2005open}. Ce protocole permet l'envoi et la réception de données à travers le réseau en UDP et TCP. Le protocole réseau TCP permet le transfert de paquets sans perte de données mais se trouve être moins efficace en temps réel. Le logiciel \textit{Max/MSP} permet une communication via le protocole UDP. 
\par Un autre élément essentiel lors de la production audiovisuelle en temps réel concerne les capacités des machines à notre disposition. Nous tenons de plus à ce que ces outils soient utilisables par des personnes possédant des ordinateurs bon marché. \textit{Jitter} offre la possibilité d'effectuer tout ou partie des calculs matriciels, soit sur le processeur principal, soit sur le processeur graphique. Cela permet entre autres d'alléger et équilibrer le processeur principal dédié à l'audio. \textit{Jitter} a aussi été désigné afin d'adapter le taux de rafraîchissement des images vidéos en fonction des ressources disponibles sur l'ordinateur~\cite{jones}.

\par En outre, Max offre de nombreuses possibilités concernant les interactions humain--machine. Premièrement, le protocole MIDI est pris charge, assurant une grande diversité de contrôleurs par les ports USB. L'objet \textit{hi} permet lui d'interagir avec de nombreuses interfaces de jeux-vidéo (trackpads, manettes, joysticks...). De plus, de nombreux \textit{externals} permettent de prendre en charges des controleurs USB tels que le \textit{Ableton Puh}, le \textit{sensel} et les \textit{Roli Blocks}~\footnote{~Ces \textit{externals} sont disponibles dans le \textit{package manager} de Max}. En effet, comme décrit précédemment, le geste et le mouvement sont essentiels pour le contrôle de paramètres multimodaux.  

\section{Discussions}

\par L'état de l'art présenté ci-avant montre l'approche interdisciplinaire nécessaire à la réalisation d'une oeuvre vidéomusicale. Ce corpus académique démontre aussi qu'aujourd'hui, peu de liens réels et directs n'ont été fait quant aux congruences créées lors de la perception d'un élément visuel et la perception d'un élément sonore. Nous savons en revanche que le cerveau humain réagit en contexte d'éléments multimodaux, utiles à la compréhension d'un évènement sensoriel. De fait, la première étape de notre travail de création constitue la création d'un corpus d'images et de sons cohérent afin de pouvoir profiter d'un premier médium congruent et concret dans la diffusion des éléments audiovisuels. 
\par Une autre étape essentielle est le développement d'outils nécessaires à l'analyse et la communication des différents paramètres d'un média à un autre que nous appellerons \textit{mapping}. Cette étape nécessite la compréhension des enjeux liés à l'analyse des images en temps réel et en temps différé, et de maîtriser les différents paramètres de contrôle de la synthèse concaténative par corpus et de la synthèse granulaire que nous venons d'étudier dans ce chapitre. \textit{CoCAVS} et \textit{CataRT} présentent une porte d'entrée essentielle à la compréhension de tels outils. C'est pourquoi nous baserons notre recherche et le développement des programmes sur ces outils.  

\par Pour finir, nous savons que l'oeuvre créée suivra une méthode de recherche--création qui explore les problématiques arts--sciences qui y sont liées. La méthodologie d'auto-analyse sera ainsi basée sur la médiologie musicale de Régis Debray et Vincent Tiffon.

Ces modules seront utilisés au cours d'une création audiovisuelle immersive~: $\epsilon\upsilon\tau\epsilon\rho\pi\eta$\footnote{~En français : Euterpe}. 
\chapter{ViVo}
\par \textit{ViVo} est un ensemble de patchs \textit{Max} open source disponibles sur GitHub\footnote{~\href{https://github.com/ircam-ismm/vivo}{https://github.com/ircam-ismm/vivo}}. Basé sur \textit{MuBu} et \textit{CataRT}, cet outil a été développé en stage de fin de Master au sein de l'équipe ISMM\footnote{~Interaction Son Musique Mouvement} de l'IRCAM. Dans ce chapitre, nous décrirons en premier lieu le développement de chaque module d'analyse vidéo. Nous verrons dans un deuxième temps comment nous avons lié les paramètres analysés par ces déscripteurs vidéo avec les paramètres de contrôles de synthèse granulaire. Nous nous concentrerons enfin sur le développement d'un environnement de VJing (\href{https://github.com/Matfayt/ViJo}{\textit{ViJo}}) dans \textit{Jitter} dans le but de modifier la vidéo diffusée et analysée en temps réel par \textit{ViVo}.

\section{Modules \textit{Jitter} pour l'analyse vidéo}

\par R. Jones et B. Neville citent \textit{Jitter} comme étant parfaitement adapté au traitement vidéo, et décrivent son fonctionnement global ainsi que les possibilités d'optimisation des ressources accessibles à l'utilisateur \cite{jones}. Si le texte date de 2005 (version 4.5 de \textit{Max}), les informations mentionnées semblent correspondre à la version actuelle de \textit{Max/MSP} (8.5.4).
\par L'analyse de paramètres vidéos pour la musique nécessite le développement et/ou l'adaptation d'outils afin de pouvoir extraire des valeurs exploitables (que ce soit en temps réel ou non). Nous avons choisi de créer quatre modules d'analyse vidéo -- basés sur l'état des connaissances qui précède -- que nous décrirons dans cette partie. Les modules d'analyse sont créés sur \textit{Jitter} (\textit{Max/MSP}) pour les raisons de compatibilité et pour les possibilités offertes par le logiciel décrites ci-avant. En effet, Mubu et CataRT sont des \textit{externals Max} et l'intégration de nouveaux modules d'analyses est facilitée lorsqu'ils sont réalisés dans l'environnement \textit{Max}. De plus, nos connaissances en informatique musicale basées sur \textit{Max} nous ont permis de nous adapter à l'environnement \textit{Jitter}. Enfin, les logiciels de VJing tel que Resolume ou de mapping tel que MadMapper ne permettent pas de programmer des outils d'analyses aussi bas-niveau que \textit{Jitter}.


\subsection{\textit{Warmness}}
\par Puisque l'étude de l'équipe Perception et Deisgn Sonore de l'IRCAM propose une analyse des propriétés physiques qui participent à la description sémantique de la chaleur d'un son~\cite{pds}, nous proposons d'analyser la chaleur des images afin de pouvoir créer des effets congruents ou incongruents lors de notre pièce immersive. Notons que le lien entre un son chaud et une image chaude n'est ici pas démontré, mais permet de proposer une congruence via la sémantique de descripteurs audio et visuels. Le but ici est de pouvoir extraire un facteur de chaleur d'une vidéo analysée en temps réel.
\par
\begin{figure}[htbp]
\centering
\includegraphics[width=0.35\textwidth]{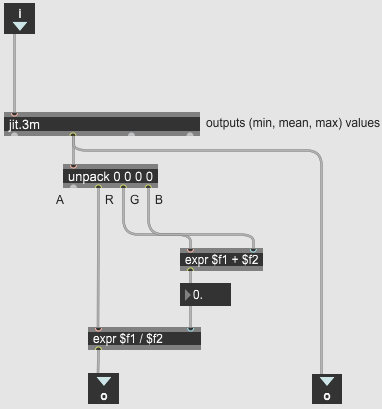}
\caption[warmness]{\label{warmness_analysis}. Le patch \textit{warmnessanalysis} premièrement intégré à ViVo.}
\end{figure}

 Une première tentative d'analyse (figure \ref{warmness_analysis}) a consisté à déterminer la chaleur de l'image via le \textit{ratio} des trois couleurs de l'espace RGB (\textit{Red, Green, Blue}). En effet, puisqu'une couleur est décrite comme chaude lorsqu'elle se trouve du côté rouge de l'espace RGB, nous avons premièrement développé un programme qui détermine si la quantité de rouge présente dans toute l'image dépasse la somme de quantité de vert et de bleu. 

\par Après avoir montré le déscripteur à différents sujets avec différentes images de synthèse, nous nous sommes aperçus que les résultats obtenus n'étaient pas ceux escomptés. Certaines images qui semblaient très froides étaient analysées par l'algorithme comme plutôt chaudes. 

\par Cela est dû à deux facteurs décrits par Michail Dimopoulos et Thomas Winler en 2014 dans un article introduisant de nouvelles techniques d'analyses d'images et de vidéos \cite{dimopoulos2014imagewarmness} basées sur la chaleur. Tout d'abord, les auteurs démontrent que les différentes couleurs perçues par les humains provoquent certaines émotions chez certains individus. Plus encore, ils soulignent que des psychologues qui analysent les effets des couleurs tendent à diviser les couleurs en deux groupes distincts~: chaud et froid. Puisque la perception de chaud ou de froid est déterminée de manière dichotomique par une seule couleur et qu'une image est constituée de plusieurs pixels, il s'agit de déterminer l'influence de chacun de ces pixels sur la perception globale de chaleur d'une image. Nous avons donc suivi cette étude afin de réaliser un \textit{patch} \textit{Jitter} qui analyse la chaleur globale d'une image. 

\par La méthode de développement s'est déroulée comme suit~: 
\begin{itemize}
    \item Prototypage des différentes fonctions décrites dans le texte sur \textit{Jitter}.
    \item Tests avec différentes sources vidéos.
    \item Création d'un script \textit{Gen}\footnote{~\textit{Gen} est une extension de \textit{Max} permettant d'effectuer des scripts dans une \textit{codebox} dans un langage de programmation proche du C.} permettant d'effectuer l'ensemble des calculs de façon optimisée.
    \item Intégration des scripts dans une abstraction \textit{Max} dédiée.
\end{itemize}

\par Premièrement, les calculs réalisés par M.~Dimopoulos et T.~Winler sont réalisés dans l'espace HSV. Puisque \textit{Jitter} utilise par défaut l'espace RGB, nous avons converti les espaces grâce à l'objet \textit{jit.colorspace}. Calculé à partir des valeurs RGB de l'image, l'espace HSV (\textit{Hue, Saturation, Value}) offre~:

\begin{itemize}
    \item Une valeur de teinte (H) qui couvre l'intégralité des couleurs de 0° à 360° sur un cercle.
    \item Une valeur de saturation (S) qui correspond à l'intensité de la couleur, sa pureté de 0\% à 100\%.
    \item Une valeur (V) qui correspond à la brillance, c'est à dire le niveau de lumière blanche qui éclaire une couleur donnée de 0\% à 100\%.
\end{itemize}

\par Puisque toutes ces valeurs sont données par \textit{Jitter} entre 0 et 255, nous les avons mises à l'échelle dans une matrice de nombres réels à virgule flottante de 0 à 1 afin de pouvoir les manipuler dans les intervalles souhaités.

\par
\begin{figure}[htbp]
\centering
\includegraphics[width=0.5\textwidth]{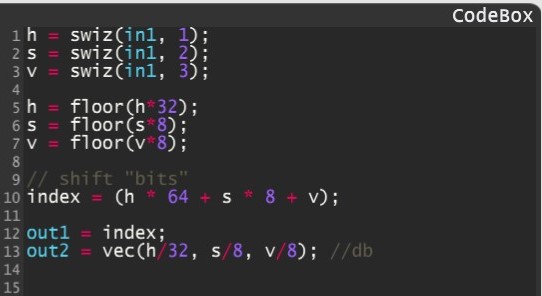}
\caption[quantize]{\label{gen_quantize} La \textit{codebox Gen} de quantization.}
\end{figure}

 La première étape à suivre consiste à réaliser un histogramme de couleurs quantifiée (figure \ref{gen_quantize}). La quantification des couleurs permet notamment de réduire le nombre de couleurs différentes présentes dans une image~: ``human perception does not differentiate between all possible color tones, but focuses on the dominant colors instead''\footnote{~<<~La perception humaine ne fait pas la différence entre toutes les nuances de couleurs possibles, mais se concentre plutôt sur les couleurs dominantes [traduction personnelle].~>>~\cite{dimopoulos2014imagewarmness}}.

\par On note par la suite l'utilisation de l'objet \textit{jit.histogram} pour réaliser l'histogramme de la matrice quantifiée. Cette étape permet d'associer un nombre de pixels à chaque valeur de couleur calculée précédemment. Cette étape pourrait être améliorée dans notre analyse de chaleur en réalisant un histogramme pour chaque image quantifiée en utilisant le \textit{median cut algorithm} suggéré par les auteurs~\cite{dimopoulos2014imagewarmness}. 

\par Par la suite, nous devons calculer la chaleur $\theta_\textit{n}$ pour chaque couleur \textit{n}. Nous devons pour cela assigner de façon binaire un facteur \textit{$T_n$}(\textit{$H_n$}) en fonction de sa valeur de \textit{Hue}~\cite{dimopoulos2014imagewarmness}:

$$
T_n(H_n) = \left\{
    \begin{array}{ll}
        -1 & \mbox{if 75° < $H_n$ < 285°} \\
        +1 & \mbox{if 0° $\leq$ $H_n$ $\leq$ 75° or 285° $\leq$ $H_n$ $\leq$ 360°}
    \end{array}
\right.
$$

\par
\begin{figure}[htbp]
\centering
\includegraphics[width=1\textwidth]{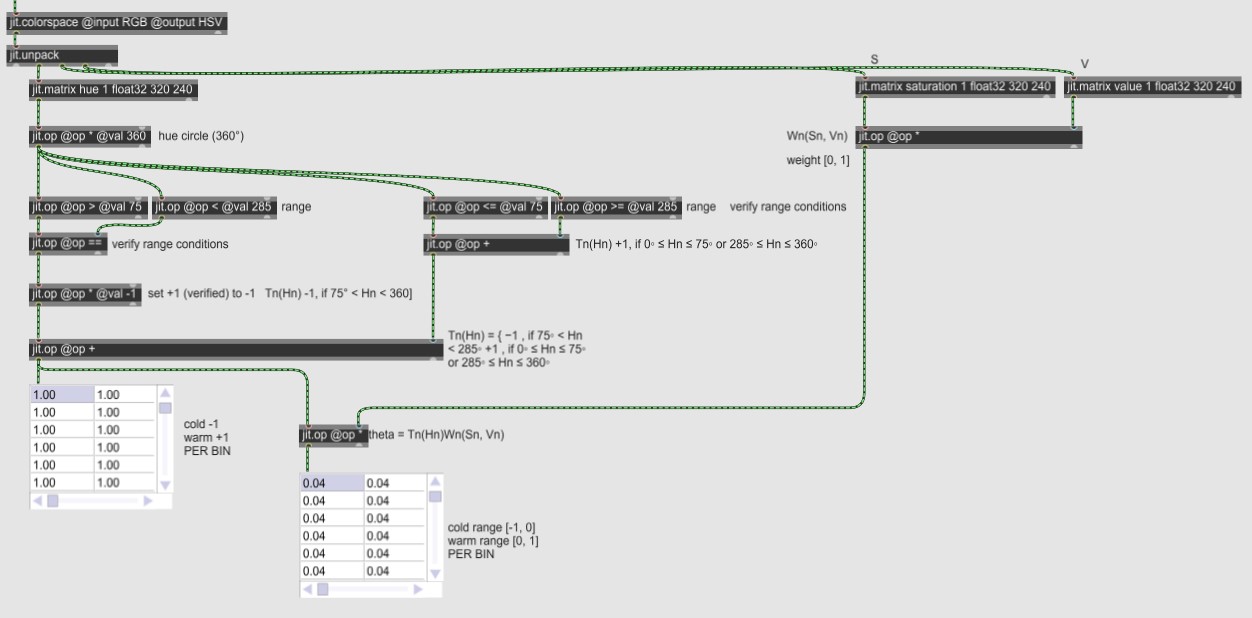}
\caption[warmnessjitter]{\label{warmness_jit} La fonction \textit{warmness} dans \textit{Jitter}.}
\end{figure}

\begin{figure}[htbp]
\centering
\includegraphics[width=0.60\textwidth]{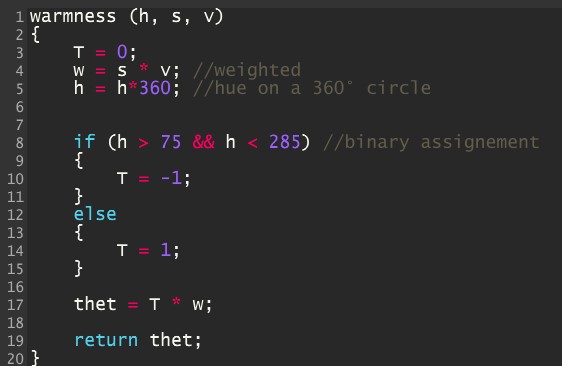}
\caption[warmgen]{\label{warm_gen} La fonction \textit{warmness} dans \textit{Gen}.}
\end{figure}

 Cette étape a premièrement été prototypée en utilisant les objets Jitter (figure \ref{warmness_jit}) puis intégrée dans le code \textit{Gen} de pré-calcul (figure \ref{warm_gen}). 

\par En outre, nous devons effectuer une fonction de pondération \textit{$w_n$}(\textit{$S_n$}, \textit{$V_n$}) afin de quantifier le niveau d'impact de chaque couleur en fonction de la saturation et de la valeur de brillance. Les auteurs proposent pour cela trois différentes manières de procéder. Nous en avons choisi une, la plus simple à implémenter dans \textit{Gen}~\cite{dimopoulos2014imagewarmness}~:

$$
w_n(S_n, Vn) = S_n \: V_n, \forall{ \: S_n, V_n} \in [0, 1] \\
$$

\par Ainsi, nous introduisons dans notre analyse les influences de la saturation et de la valeur, manquantes dans notre première tentative d'analyse de chaleur (figure \ref{warmness_analysis})~: ``A very bright and very saturated color gives a strong impression of warmness or coldness, weight, \textit{$w_n$} is approaching 1. When the color is dark or saturated there is little to no warm/cold impression, \textit{$w_n$} is approaching neutrality, 0.''\footnote{~<<~Une couleur très claire et très saturée donne une forte impression de chaleur ou de froideur, la pondération $w_n$ est proche de 1. Lorsque la couleur est sombre ou moins saturée, l'impression de chaleur ou de froideur est faible ou inexistante, $w_n$ se rapproche de la neutralité, 0 [traduction libre].~>>}\cite{dimopoulos2014imagewarmness}. Le calcul a été prototypé dans \textit{Jitter} (figure \ref{warmness_jit}) avant d'être intégré dans le code \textit{Gen} (figure \ref{warm_gen}). La dernière étape consiste à retourner le produit $\theta_n$ de la pondération \textit{$w_n$} à l'assignation binaire \textit{$T_n$} ce qui donne la fonction \textit{warmness} complète (figure \ref{warmness_jit}).

Pour finir, le calcul de la chaleur de l'image complète $\Theta$ correspond à la somme des histogrammes pondérés~\cite{dimopoulos2014imagewarmness}~:
$$
\Theta = \sum_{n=1}^{N} f_n \theta_n \\
$$

\par Puisque toutes ces valeurs sont stockées dans une matrice, nous utilisons l'objet \textit{cv.jit.sum}\footnote{~L'\textit{external cv.jit} a été développé par Jean-Marc Pelletier.} qui permet de sommer tous les éléments d'une même matrice. Ainsi, chaque image (ou échantillon de vidéo) est analysé et renvoie une valeur dans l'interval [-1, 1]. Des valeurs négatives seront donc assignées à une image froide, des valeurs positives à une image chaude, et des valeurs proches de zéro à des images avec une chaleur neutre.


\subsection{\textit{Sharpness}}
\par Le principe de ce module consiste à déterminer le niveau de netteté (en opposition au flou) d'une image. Lorsqu'une image est nette, elle laisse apparaître le bord des objects qui la constituent. Plus l'image est floue, moins les bords sont perceptibles. On appelle <<~bord~>> en traitement d'images ``a collection of the pixels whose gray value has a step or roof change, and it also refers to the part where the brightness of the image local area changes significantly''\footnote{~<<~Un ensemble des pixels dont la valeur grise varie par paliers ou par toits, et désigne également la partie où la luminosité de la zone locale de l'image change de manière significative~>>~[traduction personnelle].}\cite{WenshuoGao2010}. La détection de bord est utilisée en traitement d'images et permet de connaître la structure globale des objets. Ses applications nombreuses en font une des fonctions du traitement d'images les plus utilisées~\cite{WenshuoGao2010}. Il existe différents algorithmes de détection de bords tels que \textit{Sobel, Prewitt} et \textit{Roberts}. Ces trois algorithmes sont présents sur \textit{Jitter} ce qui permet d'avoir accès à un large choix en fonction des besoins. La détection de Sobel permet néanmoins de représenter un bord plus épais et plus brillant que ses deux homologues~\cite{WenshuoGao2010} ; c'est pourquoi nous nous sommes basés sur cet algorithme pour la détection de netteté de notre image. 

\par
\begin{figure}[htbp]
\centering
\includegraphics[width=0.35\textwidth]{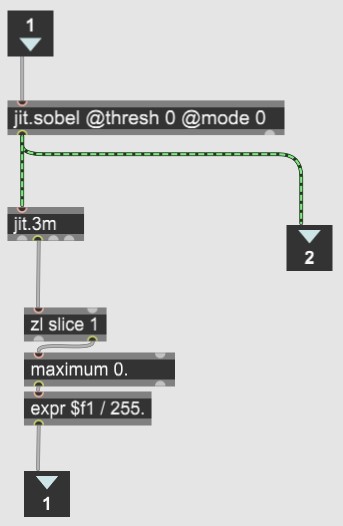}
\caption[sharpness]{\label{sharpness} Le patch de \textit{sharpness detection} dans \textit{Jitter}.}
\end{figure}

 Nous souhaitons donc après avoir effectué une détection de bords obtenir un facteur qui permet d'évaluer la quantité de bords et leur intensité dans l'image. Puisque l'objet \textit{jit.sobel} applique cette détection pour chaque couleur du spectre RGB, nous récupérons la moyenne de valeur des pixels de chaque plan avec l'objet \textit{jit.3m} et éliminons le premier élément de la liste (\textit{Alpha}) avec l'objet \textit{zl.slice}. Nous extrayons par la suite la valeur moyenne maximale parmi les trois plans RGB et divisons le nombre donné par 255 afin d'obtenir un facteur entre 0 et 1 (figure~\ref{sharpness}).

\subsection{\textit{Detail}}
\par Une analyse essentielle suggérée par N.~Collins~\cite{collins2007audiovisual} et D.~Schwarz~\cite{schwarz2023} au cours de leurs recherches sur la synthèse concaténative audiovisuelle concerne l'analyse de texture d'une image par le biais d'une transformée de Fourier. Appliquer une transformée de Fourier discrète (DFT) à une image peut aujourd'hui être utilisé pour analyser, filtrer, reconstruire et compresser des images, notamment au format JPEG~\cite{pasquini2014benford}. Cet algorithme permet de représenter une image dans le domaine fréquentiel et non spatial. Dans le domaine sonore, appliquer une transformée de Fourier permet d'analyser la richesse des composantes harmoniques qui composent un son et d'en déduire son timbre. De plus, cela permet d'obtenir une représentation visuelle de ses composantes que l'on affiche à l'aide d'un spectrogramme. En traitement d'image, la transformée de Fourier permet aussi de déterminer la richesse d'un visuel et d'en extraire ses différentes composantes. La représentation fréquentielle d'une image se fait elle aussi à l'aide d'un spectrogramme. Au même titre que dans le domaine sonore, appliquer une DFT inverse à une image dans le domaine fréquentiel permet de la représenter dans le domaine spatial. Une fois le spectrogramme réalisé, nous pouvons en analyser certaines zones afin de déterminer le niveau de basses et hautes fréquences qui composent une image. En d'autres termes, le domaine spectral permet de représenter les composantes sinusoïdales d'une image.

\par Le but de ce module et donc de déterminer le niveau de détail global d'une image, et le niveau de détail pour une plage de fréquence donnée. Nous devons pour cela réaliser des bandes de pixels du spectrogramme qui correspondront à des plages de fréquences. Par la suite, les utilisateurs pourront modifier la taille ainsi que la position de chaque bande, et donc analyser une plage de fréquence souhaitée.

\par
\begin{figure}[htbp]
\centering
\includegraphics[width=0.8\textwidth]{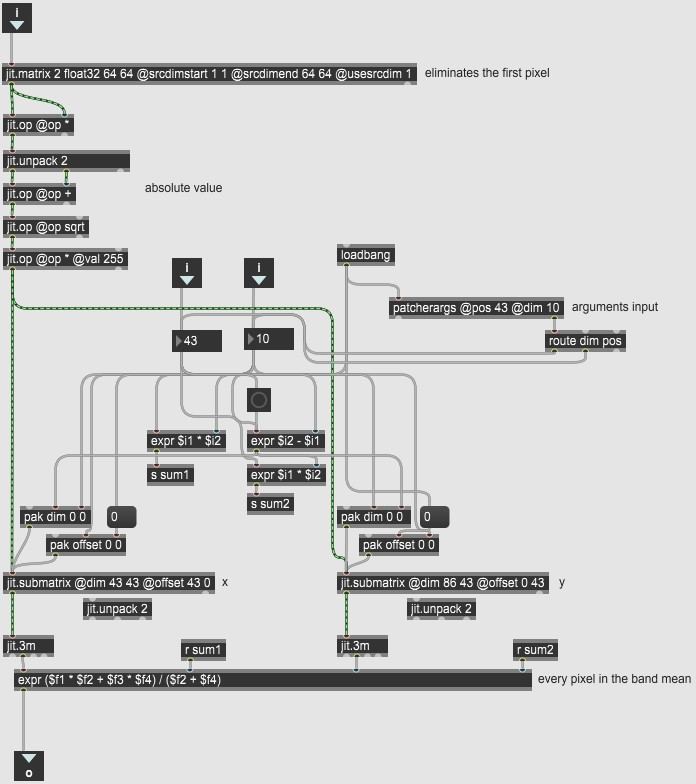}
\caption[fftbands]{\label{fftbands} L'abstraction \textit{fft\_mean\_bands} dans \textit{Jitter}.}
\end{figure}

\begin{figure}[htbp]
\centering
\includegraphics[width=1\textwidth]{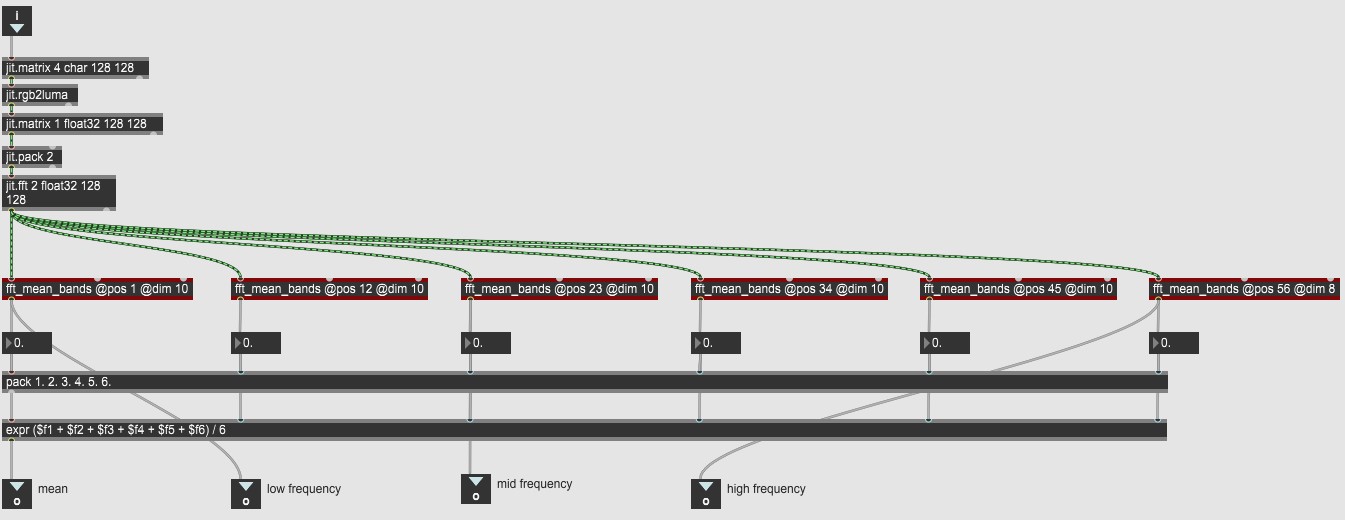}
\caption[fftduplicate]{\label{fftduplicate} Le patch \textit{detail} dans \textit{Jitter} contenant plusieurs abstractions \textit{fft\_mean\_bands}.}
\end{figure}

 Pour cela nous avons réalisé une abstraction nommée \textit{fft\_mean\_bands}\footnote{~Pour <<~moyenne d'une bande de transformée de Fourier~>>.}. Cette dernière pourra prendre deux arguments~: la position et la taille de la bande. Ainsi, après avoir réalisé la DFT avec l'objet \textit{jit.fft}, nous réalisons des bandes sur les deux axes (vertical et horizontal) à l'aide de l'objet \textit{jit.submatrix} et calculons la moyenne de valeur de tous les pixels dans la bande donnée (figure~\ref{fftbands}). L'utilisateur doit pouvoir être en mesure de déterminer la taille et la position souhaitées de la bande. En revanche, l'objet \textit{jit.submatrix} qui permet de faire référence à une région d'une matrice, n'offre que la possibilité de référencer des formes rectangulaires. La bande créée est donc en réalité constituée de deux rectangles accolés. Cela implique que lorsque l'utilisateur fait varier la taille et la position de la bande, nous devons modifier la taille et la position de deux sous-matrices. On note par ailleurs l'utilisation de l'objet \textit{patcherargs} qui permet d'attribuer des arguments à une abstraction. Cela permet une interface utilisateur simple et compréhensive pour la modification des paramètres de sous-matrices. Cette abstraction peut ensuite être dupliquée afin de réaliser différentes bandes d'une même image à différentes positions pour analyser différentes fréquences (figure~\ref{fftduplicate}).

\par Ce module d'analyse est très complémentaire avec le module \textit{sharpness}. En effet, une image peut être très détaillée sans pour autant être nette. Grâce au module d'analyse de détail, nous pouvons déterminer le niveau de détail d'une image indépendamment de sa netteté, et inversement.

\subsection{\textit{Optical Flow}}

\par Le module d'analyse de mouvement \textit{optical\_flow} permet de déterminer le niveau de mouvement global d'une image ainsi que la direction (horizontale et verticale) et l'intensité des mouvements. Nous avons pour cela adapté l'objet \textit{cv.jit.hsflow} de l'\textit{external cv.jit} développé par Jean-Marc Pelletier. Cet \textit{external} est basé sur la technique de calcul d'\textit{optical flow} de Horn-Schunk~\cite{horn1981determining}. D'après ces derniers, ``optical flow is the distribution of apparent velocities of movement of brightness patterns in an image''\footnote{~<<~Le flux optique est la distribution des vitesses apparentes de déplacement des motifs de luminosité dans une image. [traduction personnelle]~>>}~\cite{horn1981determining}. De plus, le flux optique permet de donner d'importantes informations à propos de la disposition des objets dans l'espace, et du taux de mouvement de ces objets~\cite{gibson1977analysis}. Si le flux optique est représenté chez Horn et Schunk par des matrices de vecteurs~\cite{horn1981determining}, J.-M.~Pelletier représente en sortie de son objet des valeurs d'estimation du déplacement de pixels par rapport à l'image précédente. Ainsi, nous récupérons deux matrices~:

\begin{itemize}
    \item Une matrice pour l'axe horizontal dans laquelle les valeurs négatives représentent un mouvement vers la gauche et les valeurs positives un mouvement vers la droite.
    \item Une matrice pour l'axe vertical dans laquelle les valeurs négatives représentent un mouvement vers le haut et les valeurs positives un mouvement vers le bas.
\end{itemize}

\par
\begin{figure}[htbp]
\centering
\includegraphics[width=0.5\textwidth]{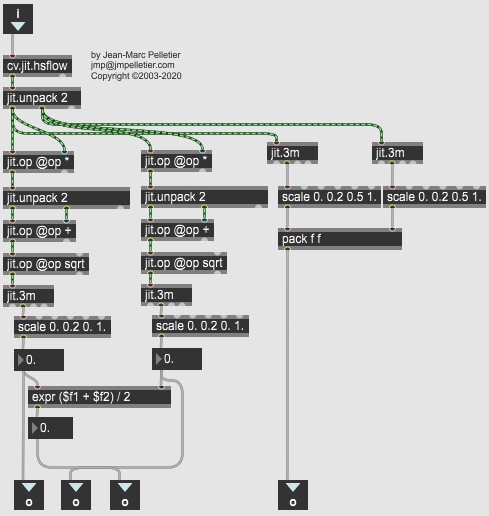}
\caption[opticalflow]{\label{opticalflow} L'abstraction \textit{optical\_flow} dans \textit{Jitter}.}
\end{figure}

\begin{figure}[htbp]
\centering
\includegraphics[width=0.5\textwidth]{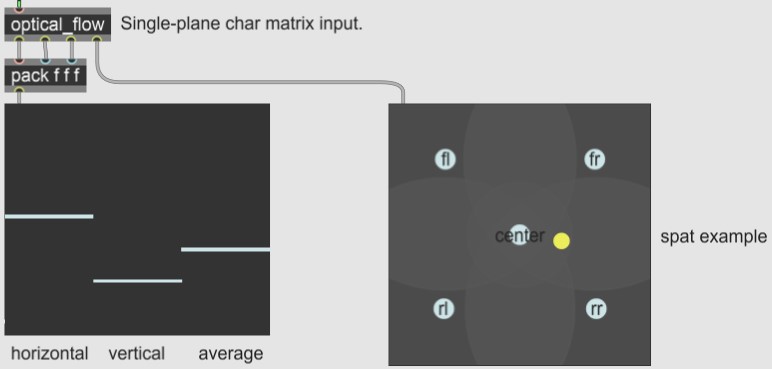}
\caption[opticalflowdata]{\label{opticalflowdata} Les données transmises par \textit{optical\_flow} dans \textit{Jitter}.}
\end{figure}

 Puisque nous souhaitons récupérer d'une part le taux de mouvement dans chaque direction, mais aussi le taux de mouvement global dans une image (indépendamment de sa direction), nous avons réadapté le \textit{patch} de visualisation de Jean-Marc Pelletier à nos besoins (figure \ref{opticalflow}). Pour chaque matrice (horizontale et verticale), nous récupérons les valeurs absolues en calculant la racine des carrés. La moyenne des valeurs de chaque pixel de la matrice permet donc d'évaluer les mouvements verticaux et horizontaux présents dans l'image. La moyenne de ces deux valeurs représente le mouvement moyen dans l'image indépendamment de la direction. De plus, nous présentons dans cette abstraction une quatrième valeur vouée à être utilisée avec l'objet \textit{nodes} de \textit{Max}. Cet objet permet d'interpoler des données de façon graphique. Ainsi, les valeurs de mouvement de l'image pourront être routées vers un algorithme de spatialisation audio, actuellement représenté avec quatre canaux (figure \ref{opticalflowdata}).

\section{Intégration et diffusion}

\par Un des principaux objectifs de \textit{ViVo} vise à diffuser l'outil afin qu'il soit utilisable par un plus grand nombre. Cela permet en premier lieu de tester l'instrument dans des conditions de création et d'en faire émerger de potentiels dysfonctionnements. La diffusion d'un instrument numérique comme celui-ci implique entre autres~:

\begin{itemize}
    \item La disponibilité sur un dépôt libre.
    \item La diffusion d'information afin de toucher une communauté (non seulement de musiciens mais aussi d'artistes vidéastes).
    \item La création de fichiers d'aide, de tutoriels et de démonstrations afin de permettre une appropriation par les individus.
\end{itemize}

\par Tout d'abord, le dépôt est disponible sur le \textit{GitHub} de l'équipe ISMM (IRCAM). Cela permet d'une part de diffuser efficacement un lien en libre d'accès, et d'autre part de faciliter les collaborations avec d'autres développeurs et artistes.

\par De plus, nous avons créé des fichiers d'aide ainsi qu'un \textit{overview}, tous organisés dans une arborescence de fichiers identique à celles des paquets \textit{Max/MSP} disponibles. 

\par
\begin{figure}[htbp]
\centering
\includegraphics[width=0.8\textwidth]{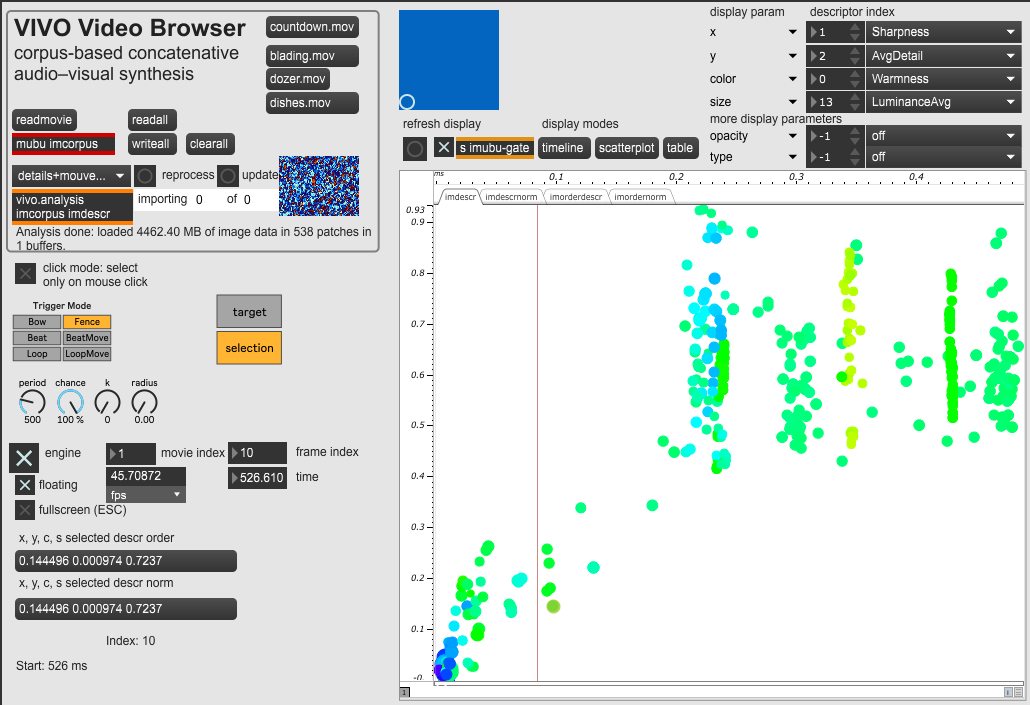}
\caption[videobrowser]{\label{videobrowser} Le \textit{patch ViVo Video Browser}.}
\end{figure}

 Afin de pouvoir intégrer les modules d'analyse à différents projets, nous les avons regroupés dans une abstraction \textit{vivo.process} qui conserve le même fonctionnement que les abstractions utilisées par \textit{CoCAVS}. Cela permet notamment de créer le \textit{ViVo Video Browser}, un \textit{patch} de synthèse concaténative visuelle qui fonctionne sur le même principe que \textit{CoCAVS}. A la place de charger un corpus d'images, l'utilisateur sera en mesure de charger une ou plusieurs vidéos. Chaque \textit{frame} sera analysée comme une image fixe par les modules développés, puis sera sélectionnable dans le nuage de points d'\textit{iMubu} (figure \ref{videobrowser}).

\par Dans notre cas, \textit{vivo.process} sera intégré dans le \textit{patch} de VJing afin d'analyser la vidéo diffusée en temps réel et de communiquer ces valeurs afin de les mapper à des paramètres de contrôle de \textit{CataRT}.

\par Ces exemples d'utilisation permettent de diffuser l'existence d'un outil ainsi que les bases de son fonctionnement. En effet, un ensemble de \textit{patchs Max} sur un dépôt (bien que libre) ne saurait être utilisé par quiconque sans la possibilité d'avoir accès à des exemples d'utilisation. Aussi, la création artistique est une forme de démonstration qui sert, par le sensible, à toucher une audience, un publique, parfois une communauté. C'est une des faiblesse de \textit{ViVo} à l'heure actuelle~: développé à l'IRCAM, au cours d'une recherche--création de Master en Acoustique et Musicologie, en somme autour d'institutions d'informatique musicale et de création sonore, il manque à ce moment précoce dans son développement encore l'entourage et l'appropriation d'une communauté d'artiste, chercheurs et développeurs vidéo. C'est cette communauté qui permettra à l'outil d'évoluer, d'être maîtrisé et utilisé sous différentes formes, bref d'être un instrument. C'est par internet, sous forme de réseau interconnecté, notamment sur les forums de discussions d'utilisateurs (\textit{e.g. Cycling'74}), que cet outil de la numérosphère trouvera son dispositif de diffusion~\cite{vincent1pour}, en tenant compte de sa communauté de pratique~: <<~Même si l'hypersphère engendre bien l'atomisation des lieux de création musicale en \textit{home-studio} indépendants et "isolés", les forums thématiques présents sur internet réintroduisent une nouvelle modalité d'individuation des plus efficaces dans la constitution de véritables communautés d'utilisateurs et de pratiques musicales.~>>~\cite{bricout2009enjeux}

\section{Mapping}
\subsection{Outils mis en place}
\par Une des utilisations principales de \textit{ViVo} consiste à établir une correspondance entre les paramètres vidéos analysés en temps réel, et les paramètres de synthèse concaténative audio. Les modules développés précédemment, permettent ainsi de contrôler de façon automatique des paramètres de contrôles de synthèse sonore. Si certains de ces modules ont été créés dans le but de contrôler des paramètres précis (\textit{e.g. warmness}), d'autres tels que \textit{detail} ont nécessité de nombreux essais afin de faire émerger des congruences pertinentes. Le but a donc été de développer des modules d'assignation adaptés au mapping à \textit{CataRT} depuis \textit{ViVo}, mais utilisables vers n'importe quel autre logiciel. Aussi, après avoir réalisé quelques essais, nous savons que \textit{ViVo} est relativement gourmand en ressources et que certaines machines peuvent ne pas être en mesure d'exécuter un programme en parallèle de celui-ci. Il faut donc pouvoir transmettre les données analysées en temps réel à une autre machine via un protocole de communication universel (de sorte à pouvoir exploiter les données hors de \textit{Max}) et qui présente peu de latence. Dans le cadre d'un jeu instrumental pulsé, les musiciens peuvent s'adapter à une certaine latence si cette dernière n'est pas aléatoire. Une latence acceptable pour une musique interactive temps réel est de l'ordre de dix millisecondes~\cite{wright2005open}. Le protocole OSC (\textit{Open Sound Control}) permet d'obtenir des latences inférieures à cette valeur. Nous devons veiller tout de même à ne pas faire transiter trop d'informations par le même canal, ce qui créerait des retards dans la chaîne de transmission d'informations.

\par Nous avons donc choisi d'utiliser le protocole OSC pris en charge par \textit{Max}. Le but est de créer des abstractions avec des interfaces visuelles qui permettent d'une part d'envoyer les données analysées par \textit{ViVo}, d'autre part de les recevoir dans un \textit{patch Max} sur la même machine, ou sur une autre machine connectée en réseau. Cette interface doit aussi permettre à l'utilisateur de mettre à l'échelle les données réceptionnées en fonction de ses besoins. En effet, nous savons que les données d'analyse sont transmises sous forme de facteur, un nombre réel à virgule flottante entre -1 et 1 pour l'analyse de \textit{warmness}, et entre 0 et 1 pour les trois autres modules d'analyse. 

\par
\begin{figure}[htbp]
\centering
\includegraphics[width=0.2\textwidth]{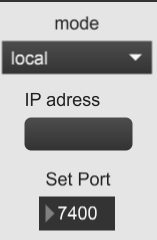}
\caption[oscsend]{\label{oscsend} L'abstraction \textit{$OSC\_send$} dans \textit{Max} en mode présentation.}
\end{figure}

 L'abstraction \textit{$OSC\_send$} (figure \ref{oscsend}) présente une interface graphique permettant de choisir si l'envoi de données est effectué en local (sur la même machine)  ou en externe (sur une autre machine) auquel cas l'utilisateur devra préciser l'adresse IP de cette dernière. Il faut ensuite préciser le port sur lequel la communication sera effectuée. Le \textit{patch} contient de plus toute la préparation nécessaire à l'envoi des données. Nous choisissons donc de respecter le modèle d'adresse OSC~: [/message valeur]. 

\par
\begin{figure}[htbp]
\centering
\includegraphics[width=0.8\textwidth]{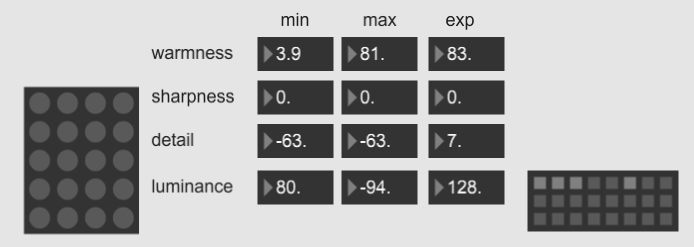}
\caption[scaler]{\label{scaler} L'abstraction \textit{scaler} dans \textit{Max}.}
\end{figure}

 L'abstraction \textit{$OSC\_receive$} permet de préciser le port d'écoute (qui a été déterminé dans l'abstraction \textit{$OSC\_send$}) sur lequel sont envoyées les valeurs d'analyse. Ces valeurs sont transmises sous forme de liste à l'abstraction \textit{scaler} (figure \ref{scaler}) dans laquelle chaque élément de la liste est séparé afin de pouvoir être traité de façon individuelle. L'utilisateur pourra par le biais de cette abstraction, déterminer la mise à l'échelle souhaitée pour chaque valeur en contrôlant le minimum et le maximum voulus. Nous avons pour cela privilégié l'objet \textit{scale} qui présente la particularité de permettre de dépasser les valeurs indiquées, ce qui n'est pas possible avec l'objet \textit{zmap} par exemple. Une fonctionnalité intéressante de l'objet \textit{scale} est de pouvoir ajouter une valeur exponentielle (ou logarithmique) à la courbe de mise à l'échelle. Nous ne sommes donc pas contraints d'appliquer un mapping linéaire entre les différents modes. Cette possibilité se trouve être particulièrement intéressante~: d'une part, la perception humaine sonore ne suit pas toujours une courbe linéaire~; d'autre part, cela permet de développer une meilleure sensibilité d'un point de vue esthétique quant au contrôle des différents paramètres de synthèse granulaire.

\par Afin de rendre le mapping multimodal modulable et interactif, nous souhaitons que les utilisateurs puissent réaliser différents mapping après réception des données. De plus, il faut pouvoir rappeler des réglages lorsque l'on change de pièce où de partie au cours d'une même pièce afin de pouvoir interagir de façon évolutive. Cela implique d'avoir une interface graphique rapidement accessible~: nous ne souhaitons pas en effet avoir à modifier un \textit{patch} au cours d'une création musicale. Nous notons donc premièrement pour cela l'utilisation de l'objet \textit{pattrstorage} qui permet de stocker les valeurs des objets dans un \textit{patch} puis de les rappeler, avec une possibilité d'interpolation entre les différents paramètres. Cet objet est notamment fait pour être utilisé en parallèle de l'objet \textit{preset} qui offre une interface utilisateur pour le rappel des paramètres. Enfin, nous proposons d'utiliser l'objet \textit{router} afin de pouvoir router les différents paramètres entrant à plusieurs différents paramètres sortants. Cet objet est mis en lien avec l'objet \textit{matrixctrl} qui permet d'offrir une interface utilisateur au \textit{router}.

\subsection{Choix des paramètres de contrôle}

\par Si le choix des paramètres de contrôle a parfois été guidé par des intuitions, nous avons tenté différentes approches de mapping une fois les corpus vidéo et sonores en notre possession. Certains de ces choix ont donc été guidés par des choix artistiques. En effet, la modification de certains paramètres ne produit pas toujours un effet suffisamment puissant pour être perceptible. Les outils précédemment développés ont entre autres permis de réaliser efficacement ces différents essais.

\par Ainsi, une des bases choisies pour la création audiovisuelle est d'utiliser les analyses \textit{warmness, detail} et \textit{sharpness}. Effectivement, si le module \textit{optical\_flow} s'avère être particulièrement efficace pour détecter des mouvements humains, ou des mouvements de caméra sur une scène relativement fixe, les essais que nous avons réalisés avec des vidéos relativement abstraites montrent que l'analyse de mouvement de l'image est très peu pertinente et fonctionnelle sur des images qui présentent beaucoup de mouvements contraires et difficilement analysables par le module. Nous avons donc décidé de ne pas l'utiliser dans le processus de création. Nous souhaitons à l'avenir pouvoir exploiter ce module, notamment pour contrôler des paramètres de spatialisation sonore, comme il le permet. D.~Schwarz a développé pour \textit{CoCAVS} un module d'analyse de brillance de l'image basé sur la moyenne de la luminance. Nous avons donc choisi de l'utiliser en plus des trois modules que nous avons développé.

\par Nous avons donc assigné les quatre paramètres comme suit~:

\begin{itemize}
    \item Le facteur \textit{warmness} sera directement lié au temps d'attaque et de relâchement de chaque grain~: plus l'image est chaude, plus les temps d'attaque et de relâchement sont longs.
    \item Le facteur \textit{detail} sera mappé à l'intervalle de randomisation de \textit{resampling}, l'attribut de \textit{mubu.concat} permettant de pitcher les grains~: plus l'image est complexe, plus l'intervalle de randomisation est grand.
    \item L'analyse de niveau de flou \textit{sharpness} est directement liée à la fréquence de déclenchement des grains~: plus l'image est floue, moins la fréquence de déclenchement est élevée.
    \item L'analyse de luminance sera reliée au facteur Q du filtre appliqué pour chaque grain~: plus l'image est lumineuse plus le facteur de filtre est élevé.
\end{itemize}

\par En effet, une première intuition nous a poussé à explorer le contrôle de la chaleur d'un grain généré par une algorithme de synthèse granulaire. Nous savons que la sensation de chaleur d'un son est notamment due au temps d'attaque de ce dernier~\cite{pds}. Nous proposons donc d'augmenter le temps d'attaque d'un grain lorsque le facteur de chaleur de l'image augmente.  Après quelques expérimentations sur des images non artificielles, nous constatons que les facteurs de chaleur générés par notre algorithme varient entre -0.3 et 0.3. En effet, nous pouvons obtenir des images plus froides ou plus chaudes en appliquant des effets qui font fortement varier la perception de chaleur d'une image. De plus, la variation d'attaque d'un grain semble plus sensible pour des valeurs basses (des temps d'attaque courts). C'est pourquoi nous avons enregistré des présélections avec un grand intervalle de temps d'attaque et avec une valeur exponentielle relativement élevée (3). De fait, nous percevons des variations sonores importantes lorsque la chaleur de l'image analysée en temps réel augmente.

\par Nous avons ensuite souhaité provoquer un fort sentiment de variation sonore en fonction de la complexité de l'image. Le module d'analyse \textit{detail} propose d'extraire une moyenne des différentes fréquences analysées dans l'image par l'intermédiaire de transformées de Fourier. Ce fort niveau de détail extrait par le module d'analyse a donc été mappé à l'intervalle de randomisation \textit{resamplingvar} de \textit{mubu.concat}. Ainsi, plus une image présente de détails dans toutes les fréquences analysées, plus le nuage de sons généré présentera de grains de hauteur tonale perceptible variée.  

\par Le module d'analyse \textit{sharpness} présente une forte complémentarité avec le module précédent. Parfois, les valeurs sont fortement corrélées et suivent une courbe similaire pour la modification d'un seul paramètre d'image. Lorsque l'on cadre une image de façon rapprochée en faisant un plan de plus en plus gros, l'image présente moins de bord (en quantité), mais aussi moins de détail notamment dans les hautes fréquences. C'est pourquoi nous souhaitons agir avec ce module sur des paramètres sonores proches du module précédent, afin de pouvoir faire émerger ces corrélations et décorrélations. Nous agissons ainsi sur la fréquence de déclenchement des grains. Lorsque cette dernière augmente, cela résulte non seulement en un nuage de grains plus dense, mais aussi en une hauteur tonale perçue plus haute.

\par Enfin, le module d'analyse de luminance propose via une moyenne du plan de luminance dans le plan HSL (\textit{Hue, Saturation, Lightness}) de pouvoir déterminer la brillance d'une image. Par un lien sémantique, nous souhaitons générer des sons plus brillants lorsque l'image est plus brillante. Comme nous savons que la notion de brillance d'un son et notamment liée à son centroïde spectral~\cite{pds}, nous souhaitons agir sur la fréquence de résonance du filtre de chaque grain. Ainsi, lorsque la brillance de l'image augmente, le fréquence de résonance du filtre augmente elle aussi, faisant augmenter le sentiment de brillance du son.

\par Si certains des contrôles sont directement mappés à la réception de valeurs d'analyse d'image en temps réel, nous avons souhaité garder un contrôle direct avec l'algorithme de synthèse granulaire. En plus de la sélection des grains dans le corpus \textit{CataRT}, nous voulons contrôler d'autres paramètres de la synthèse sonore~: \textit{mubu.concat} offre pour cela des fonctionnalités très intéressantes. En effet, en plus de pouvoir agir directement sur un paramètre de contrôle, nous pouvons agir sur sa valeur relative en fonction d'un autre paramètre. Ainsi, en plus de la valeur appliquée directement à un paramètre, nous pouvons le faire varier en fonction d'un facteur que nous déterminons. Cela implique deux choses~: premièrement, le contrôle d'un paramètre de synthèse peut modifier plusieurs paramètres de contrôle~; et deuxièmement, nous pouvons agir sur la valeur relative d'un paramètre alors que celui est modifié par les valeurs d'analyse vidéo. Ainsi, nous serons en mesure d'avoir le contrôle direct sur les paramètres de synthèse et de conserver un geste instrumental. De fait, la modification des évènements sonores pourra être sensible et ne sera pas uniquement générative et automatique. 

\section{\textit{ViJo}, un outil de VJing sur \textit{Jitter}}
\par Au cours de la pièce nous souhaitons pouvoir déclencher et contrôler en temps réel les différentes vidéos diffusées et y appliquer des effets. Cela implique de pouvoir déclencher des vidéos sélectionnées et contrôler les différents effets choisis par l'artiste vidéaste. Le VJing est une pratique artistique souvent associée à l'ajout d'événement visuel sur un élément sonore~: <<~Le VJing est une forme de performance vidéo en direct, souvent réalisée avec des DJ et des musiciens pour créer une toile de fond visuelle dans les boîtes de nuit et autres événements musicaux.~>>~\cite{taylor2009turning}. Cela implique notamment de contrôler un logiciel d'ordinateur via une interface MIDI~\cite{taylor2009turning}. Puisque les différents logiciels de VJing et mapping vidéo tels que \textit{Resolume} et \textit{MadMapper} ne proposent pas d'outil adapté à l'analyse vidéo, nous avons développés les modules d'analyse sur \textit{Max}. Cela implique donc de créer une interface de VJing pour \textit{Max}, en utilisant des outils \textit{Jitter} qui peuvent être contrôlés par des interfaces MIDI. Puisque nous sommes familiers avec des interfaces de contrôles destinées au son, et que nous sommes en leur possession, nous avons décidé d'adapter leurs protocole au VJing dans l'environnement \textit{Max}. L'espace de projection de la Fabulerie où sera réalisée la création propose une diffusion vidéo sur deux écrans et nous souhaitons pouvoir diffuser deux vidéos par écran que nous mélangerons en utilisant un curseur d'effet intégré à \textit{Max}.

\subsection{Adaptation d'un contrôleur MIDI}

\par Les contrôleurs à notre disposition sont le \textit{Ableton Push} et le \textit{Akai MPK Mini MK3}. Le \textit{push} est un contrôleur spécifiquement dédié au DAW \textit{Ableton}. Il se présente comme la vue session du logiciel et permet de déclencher des clips et des scènes au sein de pistes. Il possède aussi de nombreux encodeurs rotatifs que l'on peut assigner à des effets. C'est pourquoi nous avons conservé la logique de la vue session d'\textit{Ableton} qui nous est familière et que nous l'avons adapté à la vidéo. Ainsi, chaque piste est un dossier dans lequel chaque vidéo est contenue dans un clip. De ce fait, plusieurs vidéos provenant de différentes pistes peuvent être jouées et mixées en même temps. Les encodeurs rotatifs sont assignés à des effets appliqués à une ou plusieurs vidéos. L'\textit{external jk.push} permet d'utiliser le contrôleur \textit{Ableton Push} dans l'environnement \textit{Max}. Ainsi, tous les pads du contrôleur ont été assignés au déclenchement des différents clips vidéo et certains encodeurs ont été mappés de façon à pouvoir mixer deux vidéos sur un seul et même écran. Le contrôleur \textit{MPK Mini} utilise lui un protocole MIDI classique. Nous avons donc assigné les pads à l'activation des effets, et les encodeurs à la valeur d'application des effets avec la fonctionnalité \textit{MIDI Assign} de \textit{Max}. Ces assignations MIDI sont enregistrées dans un fichier \textit{maxmap} qui permet de garder en mémoire les assignations MIDI réalisées.

\par La conservation du geste instrumental est essentielle non seulement pour le performeur, mais aussi pour la diffusion de l'outil. Il existe d'ores et déjà une communauté d'utilisateurs du \textit{Ableton Push} et c'est grâce à ce média, cette interface (physique mais aussi conceptuelle de par l'organisation spatiale qu'elle impose) que les différents outils développés seront accessibles par un geste instrumental et formeront un instrument. C'est alors qu'instrumentiste et instrument formeront un système~: 

\begin{quote}
    <<~La fonction physique que doit remplir l'instrument de musique est précise~: Qualitative, elle est de permettre la transformation de phénomènes gestuels en phénomènes acoustiques. Quantitative, elle est de permettre, à travers cette transformation, l'émission d'une certaine énergie à destination des auditeurs à partir d'une énergie développée par l'instrumentiste. La seconde fonction ne peut être remplie qu'à la condition que la relation entre l'instrumentiste (qui est la seule source d'énergie) et l'instrument soit une interaction physique. Dès lors, toute action de l'instrumentiste sur l'instrument est indissociable d'une réaction du second sur le premier et l'ensemble instrumentiste-instrument forme en soi un système. Les phénomènes d'interaction dépendent des propriétés de l'un et de l'autre et c'est par la même, en particulier, que l'instrumentiste, à travers ces différents phénomènes, reçoit par ses sens tactiles et porprio-kinésthesiques une certaine information sur les propriétés et les comportements intime de l'instrument.~>>\cite{cadoz2001geste}
\end{quote}

\subsection{Effets}

\par L'environnement \textit{Max} propose d'utiliser \textit{Vizzie}, une collection de modules de traitement vidéo sous forme d'abstractions \textit{Jitter}. Ce modèle est particulièrement intéressant pour proposer une interface utilisateur de VJing compréhensive pour des personnes étrangères à \textit{Max}. Ainsi, chaque effet propose une interface graphique, ce qui lui permet d'être contrôlé depuis l'écran principal de l'ordinateur.

\par Dans le cadre de la proposition artistique à venir et à la demande du vidéaste S.~Fayet, nous avons constitué une chaîne d'effets numériques que l'on peut modifier par l'intermédiaire des contrôleurs cités ci-avant. L'objectif pointé est de modifier des paramètres <<basiques>> de la vidéo, sans utiliser d'effets qui résultent en de trop fortes modifications d'image~:

\begin{itemize}
    \item \textit{MIXFADR} permet de réaliser des fondus enchaînés entre deux images. Deux paramètres sont accessibles à l'utilisateur~: l'opération de fondu appliquée, et le taux de mix des deux images.
    \item \textit{BRCOSR} permet de modifier les valeurs de brillance, contraste et saturation de l'image.
    \item \textit{ZOOMR} permet de zoomer ou dézoomer sur une image. Deux paramètres sont accessibles à l'utilisateur~: le taux de zoom et le traitement appliqué à l'image, qui dépasse dans le cas d'un dézoom plus important que la taille de l'image.
    \item \textit{MOVIEFOLDR} permet de sélectionner des points horizontaux et verticaux indépendants autour desquels un effet miroir va être appliqué à l'image.
\end{itemize}

\begin{figure}[htbp]
\centering
\includegraphics[width=0.60\textwidth]{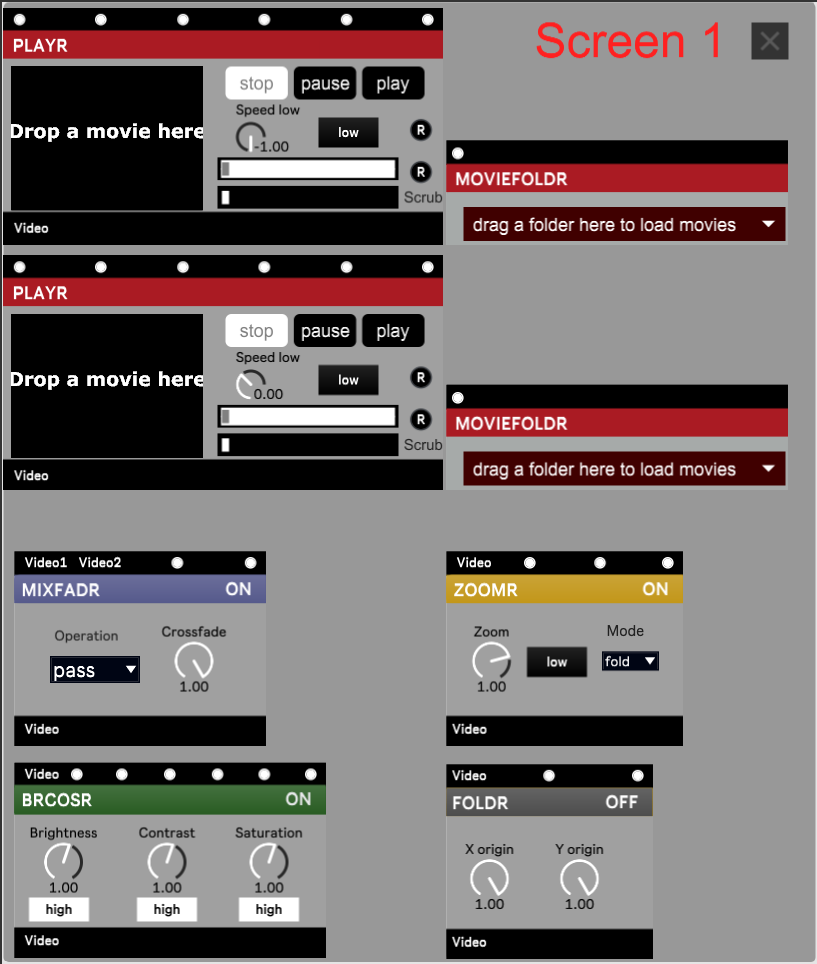}
\caption[vijo]{\label{vijo} Un des deux écrans de contrôle de l'abstraction \textit{ViJo}}
\end{figure}

\textit{Vizzie} propose aussi des modules qui permettent de stocker des dossiers vidéos (\textit{MOVIEFOLDR}) et de sélectionner la vidéo souhaitée (\textit{PLAYR}), ainsi que de projeter la vidéo dans une fenêtre flottante aux dimensions souhaitées (\textit{PROJECTR}). Cette collection de modules a donc été utilisée pour faciliter la prise en main des différentes actions à effectuer pour préparer l'installation d'une création visuelle. Ces modules ont été organisés dans une abstraction \textit{Max} qui permet de proposer une interface épurée et facile d'utilisation (figure \ref{vijo}).

\par Un des problèmes que nous avons rencontré est l'initialisation des valeurs d'effet lors de l'ouverture du \textit{patch}. En effet, pour des soucis de praticité de performance artistique, nous souhaitons que les valeurs soit initialisées dans un état neutre à l'ouverture du \textit{patch}. Pour ce faire, l'objet \textit{loadbang} permet d'envoyer des messages à des objets dès l'ouverture du \textit{patch}. En revanche, l'initialisation de l'objet \textit{jk.push}, lorsque le contrôleur se connecte, envoie les valeurs des encodeurs dans l'état enregistré par la machine. C'est pourquoi nous avons favorisé l'envoi de messages d'initialisation après que chaque encodeur du contrôleur ait été initié.

\section{Discussions}

\par L'environnement \textit{Max/MSP/Jitter} s'est avéré être particulièrement adapté pour la prise en main d'un outil qui explore les problématiques liées à la réalisation d'un instrument multimodal. Le \textit{patching} rapide, les différents outils de visualisation d'information, les format d'abstraction et de mode présentation en font un outil complet du prototypage jusqu'à la performance temps réel. En revanche, nous avons été confrontés à différents problèmes liés au principe de fonctionnement de \textit{Max} et aux limites que cela implique. La prise en main de \textit{Gen} s'est avéré particulièrement fastidieuse de par l'organisation de sa documentation et son manque de maintenance. De plus, le \textit{patching} nécessite une organisation très rigoureuse des différents objets sur une même fenêtre. Une des solutions consiste notamment à organiser les programmes en \textit{subpatchers} et en \textit{abstractions}. Cependant, cela nuit à la bonne compréhension du code de la part d'une personne extérieure ou d'une personne qui souhaite apporter des modifications à son propre code. 

\par
\begin{figure}[htbp]
\centering
\includegraphics[width=0.8\textwidth]{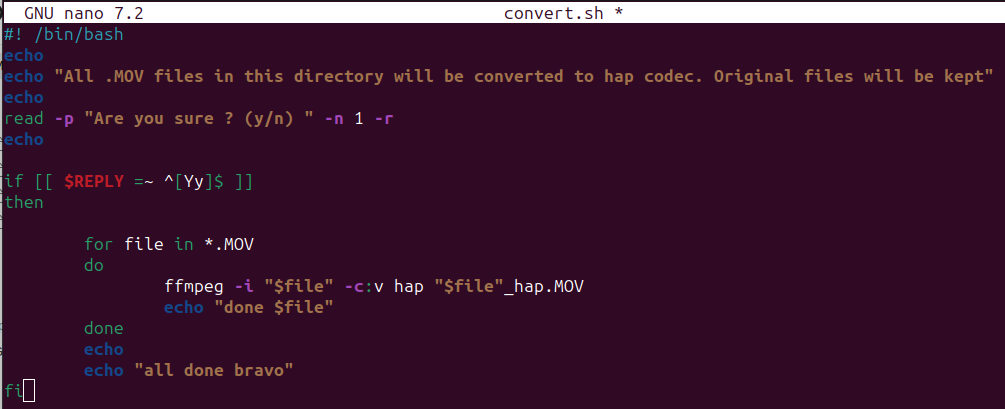}
\caption[hap]{\label{hap} Le script \textit{converter.sh}}
\end{figure}

 De plus nous avons rencontré des difficultés de lecture vidéo dans le \textit{vivo.video.browser} ainsi que sur \textit{ViJo}. En effet, les vidéos sont directement lues depuis le périphérique de stockage (disque dur), et non pas depuis la mémoire vive qui présente des capacités de lecture bien supérieures. La mémoire vive de nos ordinateurs ne permet pas d'y charger l'intégralité du corpus dessus. Nous avons donc, grâce à \textit{FFmpeg}, encodé les vidéos en \textit{hap}, un codec qui propose de meilleures capacités de lecture que le format original\footnote{~\href{https://hap.video/}{https://hap.video/}}. Le paquet \textit{FFmpeg} est accessible en ligne de commandes. Nous avons donc réalisé un script \textit{bash} disponible sur le dépôt afin d'automatiser le processus d'encodage dans un dossier vidéos (figure \ref{hap}). Cette modification s'est avérée fructueuse pour \textit{vivo.video.browser} mais pas pour \textit{ViJo}.

Enfin, les capacités de \textit{debugg} de \textit{Max} sont très limitées~: il faut utiliser pour cela la console qui s'avère parfois trop peu verbeuse. L'utilisation de points d'arrêts sur \textit{Visual Studio} par exemple, présente des capacité de \textit{debugg} bien supérieures. Le point fort de \textit{Max/MSP} réside donc pour ce type de développement d'outil dans l'efficacité de prototypage. Ainsi, nous souhaitons à l'avenir proposer de développer des outils similaires sur des environnements tels que \textit{SuperCollider} et \textit{Unreal Engine}.

\par Une fois les outils fonctionnels nous avons pu réaliser les premiers essais, dans un premier temps sur un seul vidéoprojecteur et avec une simple stéréophonie. Cela a donné lieu à un premier test de performances, sur un seul ordinateur. Les essais montrent que lorsque \textit{CataRT} fonctionne avec une latence de 15ms, nous en sommes en mesure de diffuser deux vidéos (dont une analysée par \textit{ViVo}) avec une vitesse de rafraîchissement de l'image de l'ordre de 50 fps\footnote{~Tests réalisés avec processeur \textit{AMD Ryzen 7 5000 series} et une carte graphique intégrée \textit{Nvidia GTX 1650}.}.

\par Ces répétitions ont aussi été l'opportunité de réaliser les premières expérimentations esthétiques et sensibles vers lesquels nous souhaitons nous tourner grâce à l'utilisation de cet outil. Ces intentions artistiques, vidéos et musicales, ont été développées au cours des mois de janvier à juin 2023, dans le but de présenter une pièce vidéomusicale ($\epsilon\upsilon\tau\epsilon\rho\pi\eta$), dont nous décrirons la genèse dans le chapitre suivant.
\chapter{La pièce} 

\par Le développement de \textit{ViVo} est impulsé par la volonté de réaliser une oeuvre vidéomusicale immersive collaborative (un artiste vidéaste et un artiste sonore seront impliqués). Si le développement des instruments et la création ont évolué en parallèle dans le but d'une adaptation constante à double sens, nous décrirons dans ce chapitre la genèse d'$\epsilon\upsilon\tau\epsilon\rho\pi\eta$, une pièce vidéomusicale électroacoustique immersive. Effectivement, après avoir discuté du développement technique des différents instruments, nous nous pencherons sur la démarche artistique et esthétique de notre composition. 

\par Afin de comprendre les mécanismes mis en oeuvre et les choix réalisés au cours de la création de l'oeuvre, nous étudierons en premier lieu les intentions esthétiques des deux auteurs. Cette pièce est basée sur la constitution de corpus sonores et visuels autour d'une thématique précise abordée dans la première partie de ce chapitre. Nous nous pencherons ensuite sur les enregistrements audio et visuels ainsi que les différentes techniques mises en places pour la création de ce corpus. Nous discuterons ensuite de la prise en main des instruments développés au cours de la création de la pièce, puis nous analyserons le processus créatif à l'aide de méthodes d'auto-analyse musicologiques.

\section{Esthétique}

\par La genèse de la pièce naît d'un échange oral, d'un projet commun. Après avoir évoqué l'idée de lier sons et vidéos dans une pièce immersive et de travailler la matière, nous avons commencé à réfléchir et à travailler à partir d'enregistrements sonores existants sur le thème de l'eau. Nous avons ainsi rapidement compris les possibilités offertes par la fusion de nos deux esthétiques et avons majoritairement échangé sur le sujet en utilisant un vocabulaire descriptif, sans faire référence à des artistes ou des oeuvres en particulier. La notion de collaboration artistique est inhérente à ce projet. Il apparaît comme essentiel d'intégrer au processus de création des professionnels qui maîtrisent chaque élément mis en oeuvre en son sein. A ce titre, ce chapitre met en avant le travail collaboratif qui a eu lieu au cours des mois d'octobre 2022 à juin 2023.


\par Les notes d'intention qui suivent nous ont permis de finaliser le fil d'une pensée développée au cours de 6 mois de réflexions et d'expérimentations. Cela nous a donné l'opportunité de garder un fil conducteur pour la création de la pièce. Aussi, nous pouvons désormais communiquer nos intentions au public visé afin de mieux faire comprendre les idées que nous souhaitons développer au cours de cette création. Enfin, la trace écrite d'une intention esthétique (puis d'une auto-analyse) paraît essentielle pour l'inscription de l'oeuvre dans le paysage des créations multimodales, pour sa diffusion. Les traces numériques correspondent aux paramètres enregistrés que nous pouvons rappeler sur nos \textit{patchs Max}, mais aucune trace graphique (de type partition) de notre oeuvre ne permettrait de la rejouer à l'identique (c'est que nous ne le souhaitons pas~!). Pour autant, la documentation et l'inscription d'une expérience de composition permet de contribuer à l'évolution d'un écosystème de création auquel nous appartenons. Nous avons donc rédigé une note d'intention pour chaque mode, que nous mettrons en lien par la suite. 

\subsection{Note d'intention visuelle}

Cette section est extraite d'un entretien avec Solal Fayet, vidéaste et VJ au cours du projet de développement de \textit{ViVo} et de la création $\epsilon\upsilon\tau\epsilon\rho\pi\eta$.

\begin{quote}

\par <<~Ce qui m’a intéressé quand Matéo m’a proposé de participer à ce projet il y a un peu moins d’un an, c’est le lien entre son et image en live, car cela fait partie des procédés artistiques que j’aimerais réussir à mettre en place sur mes projets personnels, et en tant qu’artiste, c’est une aubaine de pouvoir travailler avec des chercheurs pour créer un outil qui serve non seulement au développement, mais qui soit aussi prêt à l’usage pour n’importe quel artiste qui n’a pas nécessairement la main sur les outils complexes que vous développez et qui ne nécessite pas de matériel inaccessible.

\par Nous avons choisi de travailler sur le thème de l’eau pour plusieurs raisons. Tout d’abord, nous avions besoin d’un thème commun, qui puisse rendre fluide le lien entre l’image et le son, une ligne de conduite. De plus nous sommes tous deux attachés à la nature, et ma spécialité en photographie concerne les retranscriptions abstraites de la nature (ma dernière série portait sur la neige). Enfin, le lieu (Marseille) nous pousse à aller vers la mer (qui était d’ailleurs le thème de base qu’on a ensuite étendu à l’eau). J’ai passé du temps sur des bateaux spectacles ces dernières années (\textit{e.g.} l'association Bourlingue et Pacotille), sur lesquels j’avais commencé à faire des prises de sons et vidéos~: Matéo m’a rejoint dans cet univers à mon retour ce qui nous a donné l’idée de partir sur ce terrain là.

\par J’ai donc commencé à faire des vidéos il y a un peu plus de 6 mois, mais en rivière en montagne, car je n’ai pas vu la mer de l’hiver et me suis donc concentré sur ce qui m’entoure. Ce que j’ai trouvé intéressant sur notre manière de travailler, c’est que j’ai envoyé des vidéos et des prises sons au compte goûte à Matéo pendant son stage à l’IRCAM en même temps qu’il était en train de développer \textit{ViVo}, pour qu’il puisse faire des tests et développer son outil. 
Aussi, je me suis rendu compte que mon approche de la vidéo était un peu différente avec l’idée en tête qu’elle allait avoir un impact sur du son (ce qui est un peu différent que de faire de la vidéo à mettre sur du son finalement), ce qui nous à permis de beaucoup échanger à chaque retour de \textit{field recording}, sur ce que me faisait ressentir mes prises de vues, et de quelle manière elles pourraient être analysées (comme la vitesse du courant, les contrastes dûs aux reflets, les déformations visuelles, la palette de couleurs etc…). De plus, le fait de savoir de quelle manière allait être interprétée mon œuvre m’a par la suite donné de nouvelles idées de prises de vue. Par exemple, la \textit{warmness analysis} m’a poussé à plus contraster les images, voir à y ajouter des lumières artificielles. Le \textit{sharpness detection} m’a conforté dans mes prises de flous, m’a poussé à alterner flous et nets, et m’a poussé à altérer des images en utilisant l’eau comme reflet dégradant. Enfin, le \textit{optical flow} m’a poussé à avoir un corpus de vitesses de courant assez vaste et même à aller vers l’immobile.

\par En contrepoint de ma propension pour la photographie de nature, mes inspirations concernant ce projet se tournent plus vers les synthèses digitales des arts expérimentaux contemporains comme The Noise Diary\footnote{~\href{https://thenoisediary.com/}{https://thenoisediary.com/}}, ou encore Alex Guevara\footnote{~\href{https://www.alex-guevara.com/}{https://www.alex-guevara.com/}}. J’aime beaucoup ce style de visuels et notamment leur côté hasardeux et exponentiel. J’ai vraiment gardé ces images là en tête lors de mes prises vidéo et c’est l’esprit dans lequel je vais essayer de  rester durant la performance.

\par Aussi, Matéo m’a fait un \textit{patch Max} de VJing qu’il a assigné à mon \textit{Push} et mon \textit{MPK Mini}, qui sont mes instruments de musique live, de certaine matière que je fasse du VJing comme si je faisais de la musique, ce qui rend le procédé plus fluide pour moi et qui sert de liant à l’image et au son à mon sens, car j’anticipe naturellement un impact sonore lorsque je fais un mouvement.

\par Enfin, pour faire une rétrospective et critique de notre travail, je dirais que nous avons bien avancé en vue du temps que nous avons eu. Pour améliorer notre travail par la suite nous pourrions perfectionner nos \textit{piézervatifs}\footnote{~Des hydrophones dont nous décrirons le fonctionnement dans la partie suivante.}, et je dois encore expérimenter la prise de vue sous l’eau. Au niveau technique, c’est un peu compliqué de savoir car notre résidence à la Fabulerie a été très très raccourcie et nous n’avons pas vraiment pu expérimenter en conditions réelles, mais je pense qu’avec plus d'expérience et de matière nous pourrions affiner l’outil \textit{ViVo} en trouvant de nouveaux axes d’analyse, voire créer une cybernétique entre en l’image et le son (qu’ils aient une interaction réciproque).~>>

\end{quote}

\subsection{Note d'intention sonore} 

\par L'eau est un élément qui transporte l'histoire. Son cycle, par ses différents états, lui donne une place dans tous les espaces qui constituent notre monde. Sonore, visuelle et sensorielle, la pluie coule dans les rivière et se déverse, avec tous les minéraux qu'elle transporte, dans la mer et les océans. Cette roche sous forme de grains de différentes tailles se meut au grès des courants, donnant lieu à différentes matières. Le minéral n'a pas de sens propre~: c'est un élément infinitésimal de la matière inogarnique, il est désincarné. Seuls les mouvements que le corps minéral pratique et le temps qu'il traverse lui donnent la possibilité de s'incarner au sein d'une somme, d'un élément macroscopique, un nuage de grains. L'eau constitue ainsi l'élément hybride entre l'organique et l'inorganique~: comment explorer ces grains sous tous les aspects qu'ils présentent~? Comment donner à l'eau le son de la vie qu'elle contient~? Tout se passe comme si la somme des grains sonores de l'eau sonifiait sa propension à donner la vie.

\par L'approche granulaire du sonore permet dans ce but de considérer et maîtriser chaque élément sonore qui constitue un ensemble audible. 

\par La présence d'éléments sonores qui entourent la ville de Marseille dans laquelle ont été réalisés non-seulement les enregistrements, mais aussi la création artistique apparaît comme essentielle dans notre proposition. La mer est une composante indissociable de la cité phocéenne. 

\par La composition \textit{Sud} de Jean-Claude Risset~\cite{sudrisset} représente à cet égard une source d'inspiration principale dans ce projet de création. Si son corpus est constitué d'éléments enregistrés et de sons de synthèse sonore, la gestion de la perception des éléments présents et leur fusion résonne parfaitement avec notre intention esthétique. Les sons naturels enregistrés aux abords de Marseille (notamment dans les calanques) sont mis en dialogue au cours de la pièce à des sons synthétiques. Je souhaite pour ma part que les sons de synthèse proviennent tout de même de matière enregistrée, et que ces derniers soient modifiés et mis en dialogue avec des sons naturels non modifiés. \textit{Sud}~\cite{sudrisset} est une pièce qui représente une de mes introductions à la composition éléctroacoustique et par conséquent une forte source d'influence. Son approche concrète de la musique éléctroacoustique constitue un élément fondamental de mon travail de création. Les modifications apportées au son dans \textit{Sud}~\cite{sudrisset} constituent une inspiration essentielle aux effets appliquées au matériau sonore d'$\epsilon\upsilon\tau\epsilon\rho\pi\eta$~:<<~[...] J'ai fait proliférer ce matériau en combinant diverses transformations : moduler, filtrer, colorer, réverbérer, spatialiser, mixer, hybrider.~>>~\cite{brahmsrisset} 
\par Après une introduction constituée de son naturels (eau et oiseaux), J.-C.~Risset insère de subtiles sonorités synthétiques mises en dialogue avec les sons d'environnement. Dès la quatrième minute, le compositeur introduit des sons d'eau très filtrés, notamment dans les basses fréquences. Ces sonorités sont reproductibles grâce au filtre applicable à chaque grain dans \textit{MuBu}. La première approche granulaire se fait à la cinquième minute avec de sons de bois. Par la suite, à la dix-septième minute, des grains très aigus, filtrés est spatialisés font leur apparaition. C'est cet effet que l'on souhaite reproduire grâce à l'algorithme de spatialisation par grains de \textit{MuBu}. Aux dix-huitièmes et vingt-troisièmes minutes, on entraperçoit une harmonisation des sons de synthèse et des sons issus du \textit{field recording}. Les attributs de \textit{resmplingvar} contrôlés par les modules d'analsye vidéo, permettront d'obtenir un effet similaire. Effectivement, à l'écoute de l'oeuvre, j'ai pu mettre en parallèle les sonorités proposées par J.-C.~Risset avec les possibilités offertes par \textit{CataRT}. En ce propos, \textit{Sud}~\cite{sudrisset} est apparu -- en addition à la nature des éléments de \textit{field recording} --  comme une inspiration concrète à l'esthétique souhaitée d'$\epsilon\upsilon\tau\epsilon\rho\pi\eta$.

\par Une autre source d'inspiration conséquente concerne l'approche immersive de la création qu'a développé Iannis Xenakis au cours de sa carrière de compositeur et d'architecte. Sa considération de la place de l'espace et des modes dans la création musicale et visuelle apparaît comme une évidence dans le contexte de création actuel. Si nous avons déjà cité \textit{concret PH}, les \textit{Polytopes}~\cite{polytopes} (plus modernes) semblent l'aboutissement d'une démarche multimodale dans son approche de la composition musicale. En créant ses pièces dans différents lieux autour du monde, I.~Xenakis explore les particularités de chaque espace en proposant au public d'assister à des représentations immersives uniques. D'ailleurs, le compositeur évoque dès 1976 les problématiques de perception d'événements audiovisuels d'un point de vue de l'émotion~:

\begin{quote}
    <<~La profondeur des émotions au sens étymologique semble proportionnellement inverse à la variété et à la richesse des médias. Plus on s'achemine vers l'ascétisme de chaque activité artistique, plus se rétrécit le champ des valeurs absolues. D'où la contradiction~: "La profondeur de la demande artistique est proportionellement inverse à la richesse des moyens expressifs d'une époque donnée". Cette maxime nous conduit au refus de toute correspondance ou équivalence entre les expressions, par exemple de la vue et de l'ouïe, au moins au premier stade des conceptions. Le son et la lumière sont produits avec des moyens naturels étrangers entre eux~; les organes sensoriels équivalents diffèrent également. Le miracle de l'équivalence se produit derrière, bien plus loin que l'oreille ou l'oeil, dans les sphères profondes de l'esprit~>>.~\cite{xenakisarcha}
\end{quote}

\par Je pense pour cela que l'approche écologique de la perception permet de contextualiser un élément sonore implicite afin de l'intégrer dans un cadre lucide. L'immersion rendue possible par nos enregistrements, nos instruments et notre sensibilité artistique offre un renforcement de la contextualisation des différents éléments que nous souhaitons rendre sensible. La métaphore ainsi libérée par les sons abstraits se concentre de manière personnelle chez chaque individu <<~Dans les sphères profondes de l'esprit~>>.~\cite{xenakisarcha} C'est ce lien que nous souhaitons approcher en réalisant avant tout nos enregistrements audio et visuels dans un seul et même espace, puis en proposant une pièce immersive, comme l'étaient les \textit{Polytopes} de I.~Xenakis.

\subsection{Une intention esthétique commune}

\par Si les intentions esthétiques concernent deux modes et médias différents, on remarque que les intuitions qui mènent à la nature même de l'existence de ce projet suivent une ligne directrice commune. Encore une fois, nous souhaitons que les relations image/son créées lors de cette oeuvre immersive existent avant même l'utilisation de \textit{ViVo}. Cela implique une esthétique semblable et une mise en lien récurrente des vecteurs audio et visuels qui forme le fil de notre réflexion artistique. Cela implique aussi que nous soyons tous deux acteurs de la constitution de chaque corpus.

\par Le corpus sonore et le corpus visuel seront donc constitués aux abords de Marseille, majoritairement dans les calanques et sur la mer. Ce processus d'enregistrement à nécessité différentes mises en place que nous explorerons dans la partie suivante.

\section{Constitution d'un corpus sonore et visuel}
\par Un des éléments essentiels de cette création concerne la constitution de corpus sonores et visuels. Si le développement de ViVo permet une aide à la création lors de la restitution de la pièce, nous souhaitons tout d'abord créer des corpus sonores et visuels liés entre eux. Autrement dit, la première congruence que nous souhaitons créer est esthétique et sensible. Cela implique de constituer des enregistrements audio et des prises de vue d'éléments similaires ou identiques que nous pourrons diffuser dans notre pièce. Puisque le thème de l'eau est ici abordé dans nos intentions esthétiques, nous avons travaillé autour de l'enregistrement de cet élément. 

\par Les premières vidéos ont été réalisées sur des lits de rivière par S.~Fayet au cours du mois de février. Cela a, en premier lieu, permis de mettre \textit{ViVo} à l'épreuve dans un contexte de création. Effectivement, il est important que les vidéos analysées en temps réel ne ralentissent pas le taux de rafraîchissement lors de la diffusion sur vidéoprojecteur. De plus, cela a permis de mettre en exergue l'importance ou non de certains déscripteurs d'analyse vidéo. 

\par En parallèle, nous avons réalisé des enregistrements audio à l'aide d'un enregistreur \textit{Zoom} dans les mêmes endroits que les prises vidéo. Cela nous a aidé à mettre en évidence l'importance de la localisation de la prise de son et de l'orientation des microphones sur une même rivière. Par exemple, un enregistrement près d'une cascade présentera un son proche d'un bruit rose avec peu de variations. Cela résultera en un enregistrement audio très constant, avec peu de variations, et l'analyse physique des sons par \textit{Mubu} présentera un corpus de sons très similaire. En revanche, un enregistrement réalisé dans un lieu plus calme présentera un corpus sonore plus varié et interressant à explorer.

\par C'est grâce à cette étape que nous avons pu centrer la nature des enregistrements souhaités pour la création. A ces fins, nous avons pris l'initiative de constituer notre corpus lors d'une sortie bateau en mer. Cela permet premièrement d'obtenir des angles de vues variés sur un même point d'eau en changeant d'orientation et ainsi en changeant l'angle d'incidence de la lumière, mais aussi d'obtenir des prises vidéos sous-marines en s'immergeant dans l'eau à différents endroits. De plus, nous pourrons ainsi aisément obtenir les bruits générés par la coque d'un voilier en contact avec l'eau, agissant parfois en caisse de résonances. Cette sortie en mer, a été réalisée le 18 Mai 2023 dans les calanques de Port Miou, Port-Pin et En-Vau, à bord de Briséis, un voilier de 1930 protégé au titre des monuments historiques\footnote{~\href{https://www.afyt.fr/bateaux/BRISEIS}{https://www.afyt.fr/bateaux/BRISEIS}}.

\subsection{Fabrication d'hydrophones}

\par Les hydrohpones sont des microphones destinés à être utilisés sous l'eau. Souvent onéreux, ces appareils sont des transducteurs éléctroacoustiques qui permettent d'enregistrer des sons sous-marins. 

\subsubsection{Principe de fonctionnement}
Puisque nous n'en possédons pas, nous avons conçu ce que nous appelons \textit{piézervatifs}, des hydrophones bon marché destinés à être branchés sur un enregistreur portatif. Ces microphones ont été conçus pour les besoins spécifiques d'$\epsilon\upsilon\tau\epsilon\rho\pi\eta$ et s'intègrent donc dans notre processus de création artistique. Les différents problèmes rencontrés dont nous discuterons plus tard en on fait des objets spécifiques à usage unique. Leur éphémérité est d'ailleurs un élément essentiel du traitement que nous allons appliquer au corpus sonore constitué. Pour comprendre leur fonctionnement et la nature des sons enregistrés, nous devons aborder les détails techniques de leur conception.

\par Le \textit{piézervatif} est basé sur l'utilisation de microphones piézoélectriques. Ces microphones sont des transducteurs capables de générer un signal à partir de vibrations mécaniques. Après avoir soudé un câble asymétrique blindé\footnote{~Le piézo ne nécessite pas l'utilisation de câbles symétriques. Un câble asymétrique blindé est constitué de deux pôles~: la masse est entourée autour du pôle positif afin de prévenir les interférences.}, la capsule a été recouverte d'époxy afin de protéger les soudures, et d'un scotch en aluminium relié à la masse pour éviter des interférences avec le signal d'enregistrement. Par la suite, nous avons soudé l'autre extrémité du câble à un connecteur XLR mâle, nous avons introduit la capsule piézoélectrique dans un préservatif (d'où le nom "\textit{piézervatif")} que nous avons étanchéifié à l'aide de résine époxy. Si l'eau est un fluide, nous avons estimé que sa densité serait suffisante pour exciter le piézo et obtenir un signal. Nous avons réalisé un deuxième prototype qui a consisté à coller la capsule piézoélectrique dans un couvercle de bocal métallique, espérant ainsi créer une surface solide suffisante pour améliorer l'amplitude du signal généré. De plus, nous souhaitions pouvoir enregistrer les sons de la coque à l'extérieur du bateau, sous l'eau. Cela implique de devoir presser la capsule contre la coque. Si la pression était directement appliquée sur la capsule, nous obtiendrions un bruit supplémentaire, dégradant ainsi le rapport signal/bruit de nos enregistrements. 

\begin{figure}[htbp]
\centering
\includegraphics[width=0.5\textwidth]{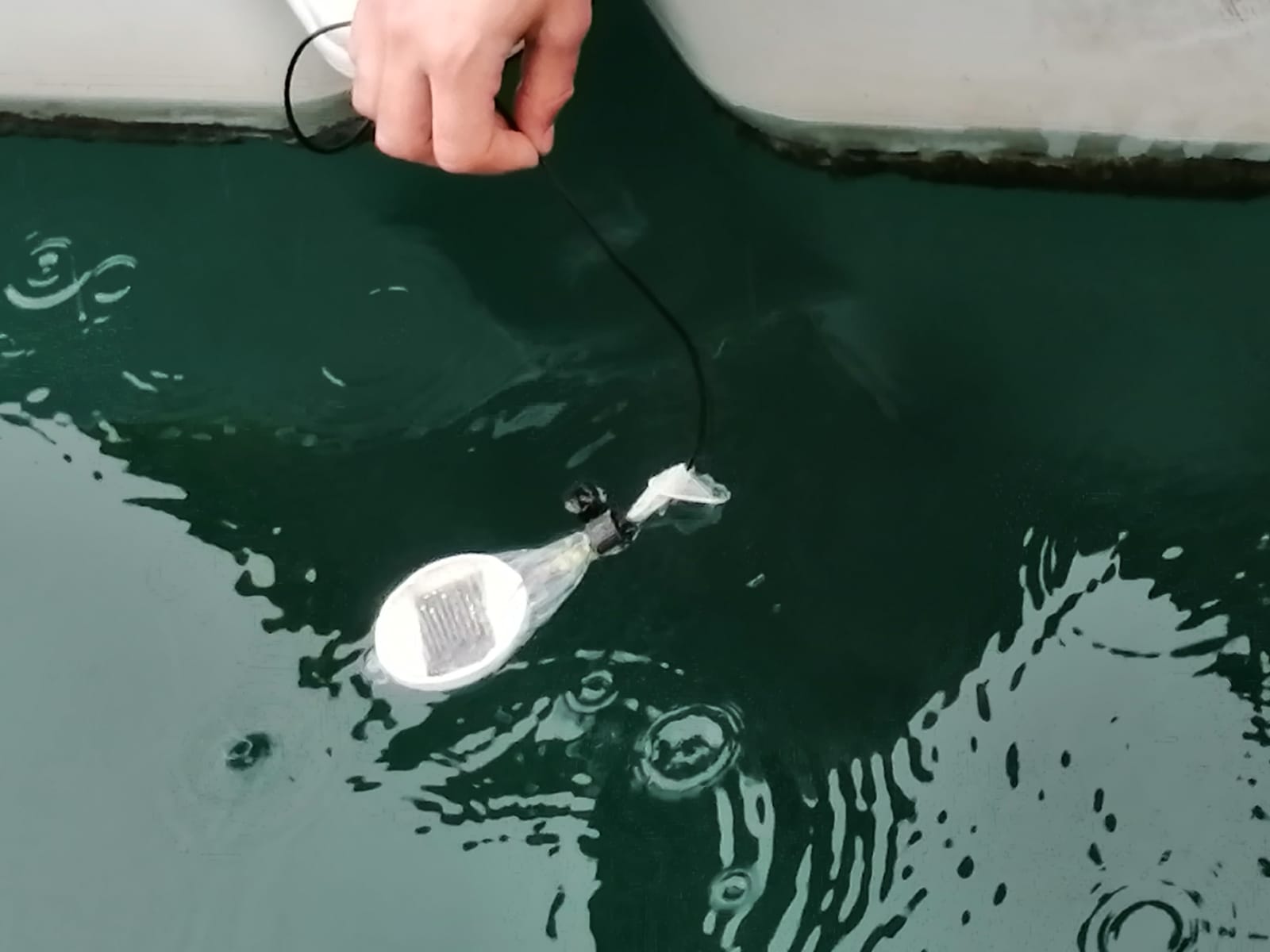}
\caption[piezervatifs]{\label{piezervatifs} Le \textit{piézervatif-couvercle} en cours d'enregistrement.}
\end{figure}

\par Dans un premier temps, les prototypes ont été testés dans un évier. Cela nous a permis en premier lieu de s'assurer de l'étanchéité et du bon fonctionnement des microphones, mais aussi de réaliser des premiers enregistrements dans le but d'affiner et de perfectionner leur utilisation \textit{in situ}. Nous avons conservé les enregistrements-tests dans le corpus de sons.

\subsubsection{Améliorations futures}

\par Tout d'abord, les capteurs piézoélectriques ont des impédances très élevées. Hors, afin d'obtenir un rapport signal/bruit acceptable, l'impédance de sortie d'un microphone doit être grandement inférieur à l'impédance d'entrée de l'enregistreur. Ils ne sont donc pas faits, à priori, pour être branchés directement à un enregistreur de type \textit{Zoom H5}. En effet, nous avons constaté lors de nos enregistrements que nous devions appliquer des gains d'entrée très élevés pour obtenir un signal compris entre -12dB et -6dB. Après écoute des prises de sons, de nombreux enregistrements se trouvent être bruités et ne sont pas utilisables. Nous souhaitons donc à l'avenir et préconisons de fabriquer des pré-amplificateurs afin de redresser l'impédance de sortie des microphones, et obtenir des enregistrements moins bruités.

\par D'autre part, le \textit{piézervatif-couvercle} (bien qu'ayant les propriétés attendues) présentait un signal filtré dans les hautes fréquences. Si l'effet n'était pas attendu, il nous a permis de réaliser des enregistrements avec un timbre différent du \textit{piézervatif} classique. De plus, l'amplification du couvercle n'a pas été aussi forte que nous nous l'attendions. Nous pensons que cela est notamment dû au fait que le couvercle était épais et non-percé. Il faudrait, pour améliorer ce prototype, percer le couvercle à un diamètre inférieur à celui de piézo avant de coller ce dernier sur le support.

\par Pour finir, l'étanchéité des prototypes a été mise à rude épreuve lors de cette journée de \textit{field recording}. Certains enregistrements ont été réalisés lorsque le bateau était en marche et la pression a permis à l'eau de s'infiltrer. Nous pensons de plus que l'eau salée altère fortement la porosité des matériaux à priori étanches. Puisque toutes nos soudures étaient protégées, cela n'a pas pollué nos prises de sons, mais a probablement altéré la durée de vie des hydrophones. Une amélioration possible consisterait à sceller l'ouverture du préservatif avec du silicone. Certains prototypes proposent de tremper les dispositifs soudés dans du goudron d'étanchéité pour bâtiment, ce qui représente une solution plus onéreuse mais plus fiable\footnote{~\href{https://felixblume.com/hydrophone/}{https://felixblume.com/hydrophone/}}.

\section{Manipulation des outils}

\begin{quote}
    <<~La synthèse du son pourrait être définie comme la composition du sonore. Avant de se risquer à approuver cette hypothèse, considérons d’abord les aspects qui entrent en jeu dans cette activité. Avant de nous pencher plus avant sur ce point, j’aimerais d’abord poser deux questions : comment le compositeur s’approprie-t-il la technologie de son époque? Comment son usage se transforme-t-il en technique~?~>>~\cite{battier2002phonographie}
\end{quote}
    
\par Ce que souligne M.~Battier dans les questions qu'il pose ici, relève de l'appropriation d'une technologie émergente au sein du processus de création musicale d'un compositeur. Effectivement, le développement d'un outil tel que \textit{ViVo}, qui permet d'augmenter le contrôle de la synthèse concaténative, nécessite une prise en main afin de pouvoir maîtriser l'outil et les différentes possibilités qu'il offre. Cette appropriation relève-t-elle d'une multiplication et d'une diversification des applications de l'outil et de l'expérience d'utilisation~? Ou bien relève-t-elle de l'orientation du développement de l'outil en fonction des besoins de la composition~?

\par Dans le cas de notre proposition d'oeuvre, le développement des outils d'informatique musicale dont il est question émane d'une part de l'augmentation d'outils existants (\textit{CataRT} et \textit{CoCAVS}) et d'autre part d'une intention esthétique et artistique dans le but de créer une oeuvre vidéomusicale immersive. Cela implique premièrement de créer les outils nécessaires dans un temps qui permet aux artistes de l'expérimenter, mais aussi de mettre à notre disposition des interfaces de contrôle adaptées à la maîtrise des outils de synthèse sonore que nous utilisons. C'est pourquoi son développement a été orienté vers les besoins de la création audiovisuelle. En revanche, certaines limites ont été imposées par les limitations de l'instrument et des ressources disponibles. Comment et avec quel corpus devons-nous réaliser nos premiers essais~? Comment allons-nous collaborer~? Quels choix faisons-nous quant au différentes possibilités de contrôle des outils de synthèse~? Toutes ces questions s'intègrent dans le processus de recherche--création dans lequel nous nous trouvons. La littérature nous a en premier lieu donné des indications sur les utilisations possibles de \textit{CataRT}. De plus, des tutoriels\footnote{~\href{https://forum.ircam.fr/projects/detail/mubu-tutorials-material/}{https://forum.ircam.fr/projects/detail/mubu-tutorials-material/}} introduisant la manipulation de \textit{MuBu} et \textit{CataRT} permettent de prendre en main les instruments et de comprendre les technologies mises en jeu.

\par La sélection et le déclenchement de grains dans l'explorateur de corpus peut être effectuée avec une souris d'ordinateur. Néanmoins, cela n'offre pas la possibilité de réaliser une sélection multiple et n'offre une navigation en deux dimensions seulement. Le \textit{Sensel Morph}\footnote{~\href{https://morph.sensel.com/}{https://morph.sensel.com/}} offre en revanche ces possibilités. En effet, cette interface USB tactile permet un contact avec plusieurs doigts. Le contact avec plusieurs doigts déclenche des flux de grains multiples de façon très naturelle~\cite{schwarz2023}. Le contrôleur permet en outre de capter la pression appliquée avec les doigts sur la surface tactile, ce qui permet de contrôler une troisième dimension. Cette dimensions peut-être mappée au gain appliqué au grain sélectionné, qui permet de conserver l'énergie de la gestuelle avec une incidence directe sur le sonore. Effectivement, dans la plupart des instruments de musique acoustiques, un geste plus énergique résulte en son plus fort (\textit{e.g.} le piano). C'est cette conservation d'un continuum énergétique~\cite{cadoz2001geste}, c'est-à-dire de l'énergie corporelle transmise par l'instrumentiste à l'instrument, que nous souhaitons explorer par l'intermédiaire de l'utilisation du \textit{Sensel Morph}~: ``This leads to the fundamental link of the performer’s gesture energy (physical pressure) to overall loudness of the audio stream''\footnote{~<<~Cela permet d'établir un lien fondamental entre l'énergie du geste de l'interprète (pression physique) et l'intensité sonore globale du flux audio. [traduction personnelle]~>>}~\cite{schwarz2023}

\par Une fois les premiers essais réalisés et les outils pris en main, nous avons été en mesure d'effectuer différentes tentatives sensibles, puis de fournir une proposition artistique que nous auto-analyserons dans la partie suivante.

\section{Auto-analyse}

\begin{quote}
    <<~La réception de l'oeuvre d'art s'effectue selon des orientations diverses parmi lesquelles deux pôles se dégagent. L'une de ces orientations porte sur la valeur cultuelle de l'oeuvre d'art et l'autre, sur sa valeur d'exposition~>>~\cite{benjamin}
\end{quote}

\par Cette partie traite de l'auto-analyse de la forme temporelle et de la diffusion de notre pièce. Nous étudierons les éléments sociaux et organisationnels qui nous ont poussés à faire certains choix quant à la diffusion d'informations et l'approche envers le public invité, mais aussi l'application des éléments évoqués dans les notes d'intentions esthétiques. Nous apporterons un regard critique sur le déroulement du processus de création ainsi que les évolutions à apporter dans le but d'améliorer et d'enrichir le fonctionnement des instruments et l'esthétique mise en oeuvre.

\subsection{Une approche médiologique}

\par Reste donc à identifier pour cette proposition de création une méthode d'analyse cohérente avec l'environnement global d'émergence de la pièce. D'après Vincent Tiffon, <<~La musicologie générale vise à interpréter les oeuvres à travers leurs traces matérielles~>>~\cite{vincent1pour}. Nous comprenons ici que l'esthétique dans laquelle se positionne notre création ne saurait être interprétée par les méthodes d'analyses musicologiques graphosphériques puisque ces dernières ne prendraient en compte que des éléments inexistants dans notre processus de création. En effet, il n'existe pour notre proposition artistique aucune partition, aucun enregistrement sur aucun support que la musicologie générale ne saurait étudier. Mais alors quelles <<~traces~>> de notre démarche sont observables et comment ces dernières (si tant est qu'elles existent) peuvent présenter un matériau analysable de façon cohérente ? Il faut extraire de notre pratique des observations accessibles à l'observateur puis reconstituer le processus d'émergence de création afin de ne pas interpréter l'oeuvre dans un cadre qui ne saurait s'extraire du champ disciplinaire de l'observateur~\cite{vincent1pour}. Régis Debray initie pour cela un <<~mode original de connaissance~>>~\cite{vincent1pour} que l'on appelle <<~médiologie musicale~>>. Quelle <<~réverbération~>> (dans le sens proposé par Leonard B. Meyer~\cite{meyer} et emprunté par V. Tiffon~\cite{vincent1pour}) comme symptôme de la contagion des idées musicales nous pousse à abandonner un système de création pour en utiliser en autre~? Une observation de l'inscription des médiums dans notre écosystème sera ainsi nécessaire à la pleine appréhension du processus de développement et de création dans laquelle nous nous plaçons.

\subsection{Évolution temporelle}

\par Le traitement des corpus sonores et visuels, tous deux réalisés à l'aide d'outils numériques, révèle l'utilisation et le détournement de certains outils à des fins artistiques. Les enregistrements abstraits du matériau visuel, par la suite traités en temps réel afin d'appliquer des effets de fortes modifications (parfois au point de ne plus identifier la matière d'origine), semble vouloir imiter le traitement appliqué aux enregistrements sonores. Ces derniers, qui sont par la suite granulés, une technique premièrement mathématiquement théorisée puis expérimentée artistiquement, ont pour but de créer un nuage sonore qui ne nous permet plus d'identifier le matériau sonore d'origine. Une fois que nous savons nos outils capables de torturer la matière, deux possibilités s'offrent à nous. La chronologie de la pièce peut en effet suivre deux logiques tout à fait opposées~: de l'inidentifiable vers l'identifiable ou de l'identifiable vers l'inidentifiable~? La compréhension des techniques de traitement que nous utilisons doit être en partie comprise afin de pouvoir être sensiblement connectés avec l'audience. Puisque les gestes appliqués sur nos contrôleurs sont abstraits mais aussi invisibles (car nous sommes dans le noir), les différentes évolutions sonores et visuelles doivent être progressives afin que l'auditeur puisse se rattacher à l'identification qu'il peut faire du matériau d'origine~: <<~En graphosphère, [...] les
musiciens sur scène produisent des sons qu’ils traduisent de leur propre lecture de la
partition ; les auditeurs entendent des sons qu’ils confirment par la vision de la source sonore – les instruments.~>>~\cite{tiffon2012public}. Effectivement, l'entrée dans la graphosphère n'a pas uniquement été provoquée par l'avènement de l'écriture et donc l'augmentation de la durabilité de la trace~: <<~[...] s'inscrire dans le temps long et non plus dans l'immédiateté~>>~\cite{vincent1pour}.  Elle a aussi été amenée par l'association de l'écoute au visuel <<~par la lecture de la partition ou par le regard sur les instrumentistes.~>>~\cite{vincent1pour}. Ici, les congruences audiovisuelles et la forme de synchrèse que représente les différentes médias présentés, forment une instrumentation de l'écoute par l'oeil. On comprend donc que si l'essentiel de l'écosystème s'inscrit en numérosphère -- de part les Matières Organisées (c'est-à-dire le développement des outils, des instruments de musique) qu'il met en jeu -- le dispositif de diffusion relève davantage de la graphopshère. Les Organisations Matérialisées représentées par le lieu de diffusion et le mode de transmission marquent la présence d'un médium graphoshpérique.

Commençant dans le noir et le silence, les écrans laissent apparaître des prises de vue abstraites de mouvement d'eau. Le sonore s'installe progressivement en parallèle, permettant à l'auditeur d'identifier le matériau projeté. Les effets appliqués à l'image et au son émergent progressivement, jusqu'à torturer la matière, la rendre méconnaissable. Le climax est marqué par le mapping total de toutes les analyses vidéos au son. Chaque détail n'est pas perceptible par l'audience, mais cela a pour effet de créer un matériau très dense et difficilement identifiable tout en créant une pléthore de mouvements, tant au niveau de l'image que du son. Cette partie marque le point culminant de la pièce mais annonce aussi la dernière partie de la pièce dans laquelle intensité sonore et lumineuse diminuent. De façon progressive, la nature du matériau d'origine redevient perceptible, avant de se fondre dans le noire et le silence. Les dernières images sont convulsives accompagnées de sons torturés aigus. Au rythme d'un battement de coeur, l'audience finit par se trouver dans l'obscurité visuelle et sonore. 

\par Une fois le choix fait de l'identifiable vers l'inidentifiable, reste à déterminer quel discours nous portons à l'audience afin de communiquer nos intentions esthétiques.


\subsection{Lien métaphorique~: le public comme musiquant}

\par Ce nonobstant, comment ne pas laisser une audience dans l'ignorance face à la transformation des éléments audiovisuels auxquels elle assiste ? Quelle quantité d'informations doit-on donner au public au regard des outils que nous utilisons ? En d'autres termes, comment conserver une sensibilité métaphorique artistique sans rendre hermétique notre lien au public, sans faire de cette pièce une démonstration technologique ésotérique ?  

\begin{quote}
    <<~Dans le domaine sonore comme pour les autres domaines, la technologie a toujours été détournée au profit des artistes, quand ce ne
sont pas les artistes eux-mêmes qui inventent de nouveaux procédés techniques pour rendre
opérationnelles leurs créations. Notre questionnement sera donc davantage celui du statut de
ces technologies contemporaines et des dispositifs qu'elles mettent en oeuvre. Sont-elles capables d’offrir à nos contemporains les conditions d’une appropriation effective des
œuvres~? En d’autres termes, les technologies utilisées dans les arts contemporains sont-
elles en mesure de créer un "milieu associé" à même de faire émerger des "communautés
sensibles"~? Comment créer un public, non plus appréhendé en termes de parts de marché,
mais en termes de transmission~?~>>~\cite{tiffon2012public}
\end{quote}

\par Autrement dit, nous ne souhaitons pas mener à une perte de participation active de la part du public. Si notre proposition artistique n'est pas interactive (mais tend à proposer des outils qui peuvent la rendre), comment rendre le lien à l'audience plus interactif~? Comment rendre le public plus <<~musiquant~>>~?~\cite{vinet2005nouveaux}

\par Effectivement, \textit{ViVo} n'a pas pour but direct de créer des congruences cognitives multimodales automatisées, mais plutôt de faciliter la communication entre artistes qui modèlent chaque média dans le but d'une unification. Nous avons par ailleurs déjà évoqué la nécessité de garder un geste instrumental et un contrôle humain sur les paramètre de synthèse.

Ce rouage qu'impose la complexité des éléments travaillés doit-être graissé bien avant la pièce, au moment de notre invitation du public, par le biais de la communication. Notre premier pas vers un lien au public nécessite la diffusion d'une description et d'une affiche pour l'invitation à assister à $\epsilon\upsilon\tau\epsilon\rho\pi\eta$.

\subsection{Communication de l'évènement}

\begin{figure}[htbp]
\centering
\includegraphics[width=0.6\textwidth]{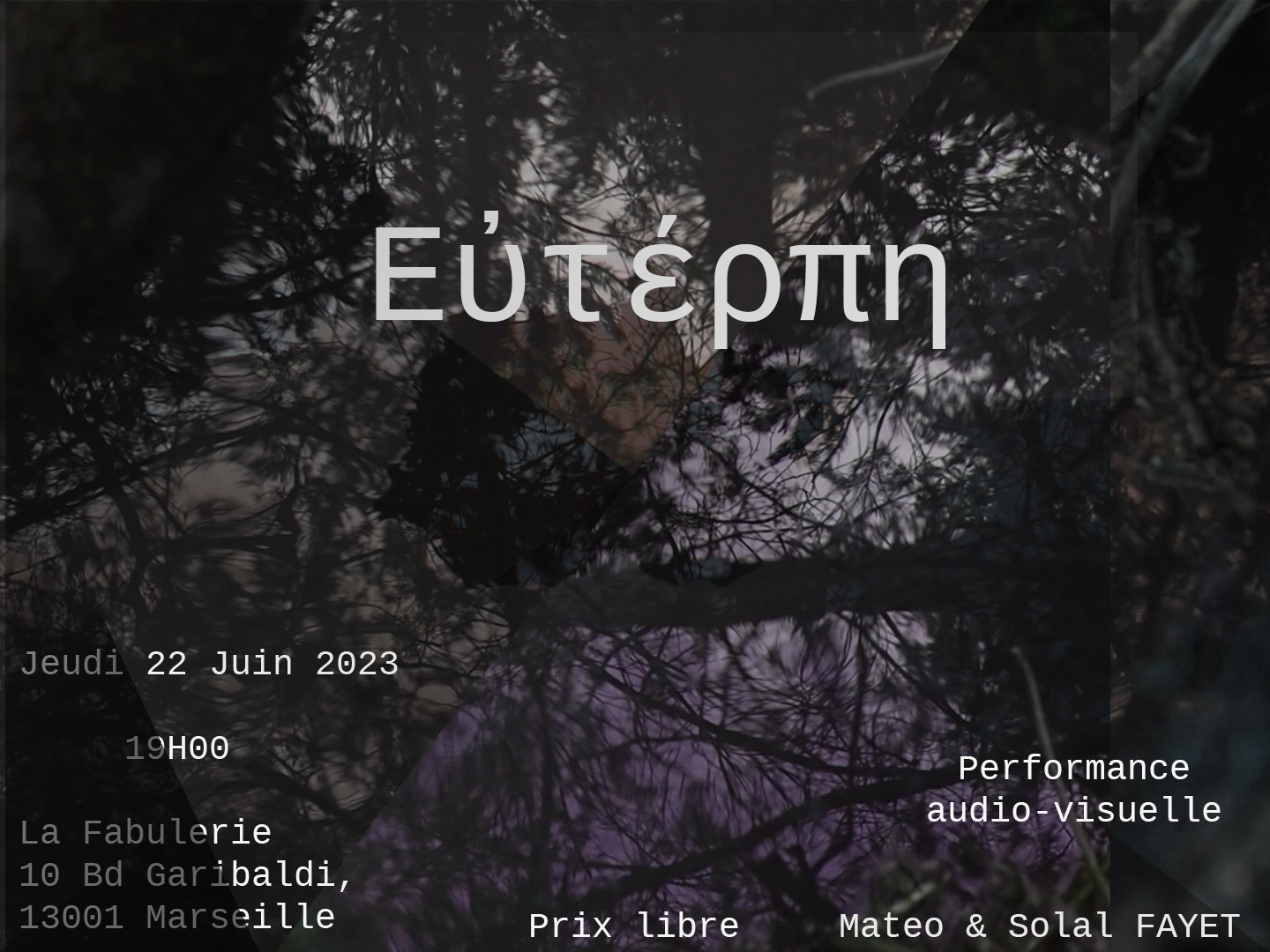}
\caption[affiche]{\label{affiche} L'affiche d'invitation à assister à la pièce.}
\end{figure}

\begin{quote}
    <<~Muse de la musique, elle nous rappelle la place de la musique dans le quadrivium
des sept arts libéraux. Euterpe, c’est une œuvre immersive dans laquelle le lien entre
l’image et le son est créé par la machine et par l’humain, dans le but de vous
emmener voyager avec nous dans un océan de mélange des sens. Immergés sous,
dans, et sur l’eau, nous vous invitons à explorer le corpus d’images et de sons que
nous retravaillerons en live pour le plus grand bien de vos haliotides et de vos
mirettes. Ce sera l’occasion de vous présenter les outils que nous développons
depuis mars 2023.
ViVo est un ensemble de patchs Max open source disponibles sur GitHub développés au cours d’une recherche-développement dans le cadre de la formation Acoustique et Musicologie d’Aix-Marseille Université avec l’aide de chercheurs des laboratoires PRISM et IRCAM.~>>\footnote{~Le texte diffusé avec l'affiche.}
\end{quote}

\par Cette affiche présente à première vue les informations essentielles au rendez-vous proposé (date, lieu et prix). Certains éléments laissent tout de même transparaître quelques unes de nos intentions. En premier lieu, le motif de fond, relativement discret est constitué de captures des vidéos qui forment le corpus visuel de la pièce. C'est grâce au reflet des arbres que l'on arrive à identifier la matière liquide qui la constituent. C'est d'ailleurs une des explorations travaillées au cours de la pièce. De plus, l'image a subi quelques traitements~: elle est composée de deux images pliées sous forme de triptique. Chaque triangle est identifiable par la légère coloration qui y est appliquée. Cela laisse transparaître la présence d'images traitées, non-naturelles. 

\par Le seul indice de présence sonore est dans la descriptions du type de pièce sur le côté droit de l'image~: <<~Performance audio-visuelle~>>. Discutons la présence d'un tiret entre les termes <<~audio~>> et <<~visuelle~>>. Cette variante ne marque à priori pas de différence avec l'écriture du terme d'origine <<~audiovisuelle~>>. Ce détail (auquel nous n'avons prêté que peu d'attention au moment de l'écrire) présente la performance et l'intention comme distinguant les deux médias. Lorsque l'on prend compte des éléments rédigés ci-avant, on comprend bien que cette maladresse ne reflète pas notre intention esthétique qui vise plutôt à envisager ce complexe comme <<~un tout~>>. Pourrions-nous ainsi varier les écritures d'un terme pour laisser sous-entendre la nature et la force du lien que nous souhaitons tisser entre nos modes~? 

\par Un autre élément essentiel à la bonne compréhension d'une oeuvre et aux évolutions que les auteurs souhaitent apporter concerne le retour de l'audience et de la communauté. Cette critique essentielle ne peut faire partie de cet écrit en raison de sa temporalité\footnote{~Ce mémoire est soumis au jury dix jours avant la création de la pièce.}. C'est une des raisons de la nécessité de communiquer l'existence d'une création audiovisuelle. Après ces retours, nous pourrons premièrement corriger des problématiques liées à la communication de l'évènement et la compréhension du public. De plus, nous serons en mesure de questionner certains choix réalisés, notamment sur la convenance du lieu, mais aussi les outils utilisés, l'esthétique de la pièce. Bref, les critiques apportées à la suite de cette création donnerons lieu à de riches évolutions, essentielles au cheminement du développement artistique mené.

\section{Discussions}

\par En définitive, $\epsilon\upsilon\tau\epsilon\rho\pi\eta$ constitue une multitude d'éléments qui lui permettent de  s'inscrire, via les différents médiums qu'elle met en oeuvre, dans différentes médiasphères~: principalement la numérosphère et la graphosphère. Une analyse plus importante devra être menée a posteriori. Effectivement, l'auto-analyse -- a priori -- exposée ci-avant présente des biais de subjectivité non-négligeables. Cette proposition artistique doit donc faire l'objet de différentes critiques objectives, tant d'un point de vue méthodologique qu'esthétique. Nonobstant une approche quelque peu antagoniste à la musicologie graphosphérique, nous nous devons de ne pas nier que les bases de nos techniques de composition et d'analyse nous proviennent de la graphosphère. On remarque d'un point de vue méthodologique que nous nous sommes tournés vers une approche écologique de la perception, considérant l'importance de la place du contexte dans la perception d'un événement sonore. Effectivement, le but premier de \textit{ViVo} était de renforcer la congruence d'événements perçus dans deux modes différents. Finalement, nous considérons cet outil comme un instrument de musique à part entière. Effectivement, les intentions esthétiques visent plus à renforcer le lien entre deux modes de créations en temps réel, pour augmenter les capacités des artistes, qu'à créer un fort effet de synchrèse chez l'auditeur. De plus, nous n'avons pas proposé à l'heure actuelle de solution pour rendre le public interactif au cours de la pièce. 

\par Au cours de cette recherche, les différents éléments scientifiques présentés montrent que les liens entre l'image et le son effectués au cours de cette création ne sont pas perceptivement avérés. Néanmoins, les intentions de création nous ont permis d'effectuer des liens cohérents dans notre proposition artistique. C'est en effet après voir développé les modules d'analyse que nous avons souhaités nous tourner vers l'élaboration de congruences sensibles, plus que scientifiquement fondées. Ce changement de cap, bien qu'effectué, est arrivé tard dans notre processus de recherche. C'est aussi une des raisons pour lesquels les modules d'analyse ont été réfléchis autour de fondations scientifiques. Aujourd'hui, nous serions en mesure de proposer une approche différente qui met en relation la recherche et la création de manière plus équilibrée. 

\par Nous remarquons aussi que nous inscrivons notre démarche artistique tant dans la \textit{visual music} que dans la vidéomusique. Si les différences entre ces deux courants ont été décrits au cours du premier chapitre, nous ne sommes toujours pas en mesure de déterminer auquel de ces deux mouvements la pièce appartient (si ce n'est aux deux). Effectivement, la conception de la mise en relation des différents modes (synchrèse) dans $\epsilon\upsilon\tau\epsilon\rho\pi\eta$ s'inscrit a priori dans une démarche proche de la \textit{visual music}. En revanche, la nature des éléments sonores et visuels présentés s'apparente davantage aux oeuvres de vidéomusiques. D'une part, cela peut provenir des moyens techniques à notre disposition (plus proches de la vidéomusique que la \textit{visual music}). D'autre part, cela révèle une forte source d'inspiration de la part des deux genres. De plus amples discussions seront nécessaires afin de déterminer le courant artistique dans lequel s'intègre $\epsilon\upsilon\tau\epsilon\rho\pi\eta$.

\par Aussi, nous avons renforcé le postulat que la création artistique est une démarche essentielle à la lutherie numérique. D'une part, cela constitue pour les artistes une mise à l'épreuve des outils développés et de l'esthétique souhaitée. D'autre part, elle permet de diffuser l'existence et le fonctionnement de différents outils afin de regrouper une communauté d'utilisateurs, ce qui apparaît comme fondamental lors du développement d'un instrument numérique tel que \textit{ViVo}. 

\par Les difficultés rencontrées au cours de la composition soulèvent les problématiques d'une part liées à l'instrument de musique et d'autre part à l'appropriation que nous nous en sommes faits. Effectivement, nous manquons d'éxperience en ce qui concerne l'utilisation de \textit{Max} et les limites du logiciel. De plus, les délais imposés nous ont poussé à faire certains choix et à ne pas utiliser toutes les possibiltiés offertes par \textit{CataRT} notamment.

\textit{ViVo}, doit être expérimenté dans d'autres contextes de création musicale dans le but de pouvoir y appliquer les modifications nécessaires et d'en faire un instrument de musique à part entière.

\par $\epsilon\upsilon\tau\epsilon\rho\pi\eta$ dans la mythologie grecque représentait la muse de la musique. Boèce définit plus tard dans les arts libéraux un \textit{quadrivium} constitué de quatre disciplines scientifiques~: l'arithmétique, l'astronomie, la musique et la géométrie. La musique, toujours représentée par Euterpe trouve alors sa place au sein d'un complexe de méthodes qui se réfère à la maîtrise des sciences. Aujourd'hui, peut-on considérer la musique comme médium d'un \textit{trivium} plus moderne que l'on appelle art, science, technologie~?

\chapter*{Conclusion}
\addcontentsline{toc}{chapter}{Conclusion} \markboth{CONCLUSION}{}

\section*{Conclusion générale}

\par Les motivations de ce projet sont nées d'inspirations artistiques et scientifiques multiples que nous avons abordées tout au long de ce mémoire. Les différentes problématiques rencontrées, loin d'avoir été un frein dans ce projet de recherche--création, ont permis d'explorer les problématiques art, science, technologie liées au développement de technologies émergentes et de créations artistiques associées. Le travail scientifique a été rendu possible par l'accompagnement de deux laboratoires de Sciences et Technologies (IRCAM) et de Musicologie (PRISM) et a été complété par l'intégration d'un tiers-lieu artistique au processus de création. Cette effervescence a enrichi le projet de manière conséquente, permettant d'étudier une grande partie des problématiques que nous souhaitions aborder au commencement de ce travail de recherche. Ce projet a notamment mis en évidence les obstacle liés au choix des technologies utilisées pour le développement de l'instrument. \textit{Max/MSP} s'est avéré particulièrement adapté pour la phase de prototypage et la prise en main des outils grâce à l'interface graphique qu'il propose. C'est aussi un outil mondialement utilisé par les artistes sonores. Néanmoins, les capacités offertes par le traitement d'images en temps réel, et le manque d'une communauté de vidéaste utilisant le logiciel en font un frein à son bon développement.

\par Par conséquent, il nous est cher de proposer des perspectives adéquates et réalisables, avec pour objectif de proposer des améliorations technologiques et créatives  concrètes pour renforcer et multiplier les possibilités artistiques multimodales, immersives et interactives.

\section*{Perspectives}

\par Au vu des outils développés, de la création audiovisuelle et des problèmes qui en émergent, nous proposons des améliorations et des évolutions aux différentes problématiques abordées dans cette recherche--création/développement.

\par D'un point de vue de la recherche fondamentale, nous envisageons de réaliser une expérience mettant en jeu les congruences cognitives multimodales de la perception humaine. Nous pourrions ainsi proposer un protocole expérimental qui vise à étudier les corrélations perceptives d'événements visuels et sonores. En liant les expériences de descripteurs sémantiques d'une part pour l'image~\cite{dimopoulos2014imagewarmness} et d'autre part pour le son~\cite{pds}, nous pourrions étudier les congruences multimodales de la perception humaine et déterminer les liens sémantiques entre les différents modes sensoriels. Cette étude permettrait de proposer des mappings adaptés dans le cadre de la création artistique, mais aussi du design. Puisque nous sommes en mesure de quantifier les caractéristiques physiques qui déterminent la qualité perceptive d'un son et d'une image, nous serions à même de mettre en place une expérience qui implique des méthodes de corrélation inverse. Cela aurait pour but de mettre en exergue les corrélations entre les comportements des deux modes de perceptions. Ainsi, nous pourrions déterminer quels facteurs au sein d'un mode ont une influence sur un autre.

\par Concernant le développement, nous souhaitons premièrement proposer un mapping multimodal du son vers l'image. Nous pourrions ainsi générer des textures et modifier des images en fonction de paramètres sonores analysés en temps réel. Cela offrirait la possibilité de créer des convergences multimodales dans les deux directions, et d'augmenter les capacités offertes par nos instruments. De plus,  nous souhaitons intégrer un troisième mode par le biais d'un retour haptique (vibro-tactile). L'objectif est de concevoir un instrument autonome basé sur une \textit{Raspberry Pi}, dans laquelle nous pourrions intégrer un explorateur de corpus accessible via un écran tactile. Cet écran générerait des textures et des vibrations en lien avec le son généré créant ainsi un complexe trimodal. L'entreprise \textit{Hap2U} propose avec le \textit{Xplore Touch} des solutions matérielles innovantes pour l'interaction multimodale en intégrant des actuateurs piézoéléctriques à des écrans tactiles\footnote{~\href{https://www.hap2u.net/haptic-technology/}{https://www.hap2u.net/haptic-technology/}}. Ces prototypes sont basées sur des \textit{Raspberry Pi} et une API basée sur le langage Python est disponible afin de rendre le dispositif adaptable. Le développement d'un tel instrument vise à explorer les problématiques liées à l'appropriation d'instruments de musique par les personnes porteuses de handicap notamment\footnote{~\href{https://ismm.ircam.fr/tag/dirti/}{https://ismm.ircam.fr/tag/dirti/}}. 

\par Enfin, nous souhaitons étendre la création $\epsilon\upsilon\tau\epsilon\rho\pi\eta$ aux arts vivants. Nous avons en effet démontré l'importance du contrôle de la synthèse par le mouvement dans le cadre de lutheries électroniques. \textit{CataRT} présente à ce titre des possibilités très adaptées. Effectivement, le corpus est exploitable dans un espace à trois dimensions et les paramètres de contrôle de la synthèse peuvent aussi être maîtrisés par les mouvements du corps humain.
La proposition artistique consiste à projeter la vidéo sur un voile semi-opaque derrière lequel se trouve un danseur équipé de capteurs de mouvements\footnote{~\href{https://ismm.ircam.fr/riot/}{https://ismm.ircam.fr/riot/}}. Ainsi, la vidéo diffusée sera visible par le public, mais aussi par le danseur. La gestuelle captée sera transmise aux moteurs audio et graphiques. Ainsi, les trois performeurs pourront interagir avec les différents modes par le biais de données numériques, mais aussi par l'intermédiaire de leur sensibilité. Nous souhaitons expérimenter ce prototype avec l'application \textit{COMO} développée par l'équipe ISMM, puis intégrer le contrôle de \textit{MuBu} par des \textit{Riot} via la \textit{toolbox PiPo}\footnote{~\href{https://ismm.ircam.fr/pipo/}{https://ismm.ircam.fr/pipo/}}.

\section*{Apports personnels}

\par Si cette recherche est pour moi l'aboutissement d'un cursus universitaire et musical, elle représente aussi la richesse des possibilités offertes par l'informatique musicale et les perspectives artistiques qui y sont liées. Ces mois de recherche et développement m'ont offert l'opportunité de travailler aux côtés de chercheurs et d'artistes dont les compétences techniques mais aussi humaines ont permis à ce projet de voir le jour. Je ne peux qu'envisager avec enthousiasme et joie les projets auxquels nous aspirons, dans l'espoir de provoquer chez d'autres ce que les mots seuls ne sauraient faire.


\bibliographystyle{plain-fr}
\bibliography{bib.bib}

\end{document}